\documentclass[%
 aip,
 amsmath,amssymb,
 reprint,%
]{revtex4-1}

\usepackage{graphicx}
\usepackage{dcolumn}
\usepackage{bm}
\usepackage[utf8]{inputenc}
\usepackage[T1]{fontenc}
\usepackage{mathptmx}
\usepackage{etoolbox}
\usepackage{xcolor}

\makeatletter
\def\@email#1#2{%
 \endgroup
 \patchcmd{\titleblock@produce}
  {\frontmatter@RRAPformat}
  {\frontmatter@RRAPformat{\produce@RRAP{*#1\href{mailto:#2}{#2}}}\frontmatter@RRAPformat}
  {}{}
}%
\makeatother

\usepackage{mathtools}
\usepackage{custom_commands}

\begin{document}
\preprint{AIP/123-QED}

\title{Continuous Floquet Theory in Solid-State NMR}
\author{Matías Chávez}
\author{Matthias Ernst}
\email{maer@ethz.ch}
\email{chavezm@ethz.ch}
\affiliation{\small Department of Chemistry and Applied Biosciences, ETH Zürich, Vladimir-Prelog-Weg 2, 8093 Zürich, Switzerland}

\date{\today}
             
\begin{abstract}
This article presents the application of continuous Floquet theory in solid-state NMR. Continuous Floquet theory extends traditional Floquet theory to non-continuous Hamiltonians, enabling the description of observable effects not fully captured by traditional Floquet theory due to its requirement for a periodic Hamiltonian.
We present closed-form expressions for computing first and second-order effective Hamiltonians, streamlining integration with traditional Floquet theory and facilitating application in NMR experiments featuring multiple modulation frequencies.
Subsequently, we show examples of the practical application of Continuous Floquet theory by investigating several solid-state NMR experiments. These examples illustrate the importance of the duration of the pulse scheme regarding the width of the resonance conditions and the near-resonance behavior.
\end{abstract}
\maketitle

\section{Introduction\label{sec:introduction}}
The ability to analyze the time evolution of the density operator under time-dependent Hamiltonians is crucial for understanding and designing experiments in solid-state NMR \cite{Haeberlen1976, Mehring:1983wm, Schmidt-Spiess}. In solid-state NMR, time-dependent Hamiltonians arise from mechanical sample rotation and the application of radio-frequency (rf) irradiation. Sample-rotation techniques like magic-angle spinning (MAS) \cite{Andrew:1958we, Lowe:1959ur}, dynamic-angle spinning (DAS)\cite{Mueller:1990wf}, or double rotation (DOR) \cite{Samoson:1988fm} are employed to average anisotropic interactions and achieve high-resolution spectra. Radio-frequency irradiation is used to manipulate the spin part of the Hamiltonian, either further eliminating or reintroducing certain terms that are averaged by sample rotation.

In solid-state NMR under MAS, multiple time dependencies are common and play a significant role in many experiments. In the simplest case of MAS without any additional rf irradiation, only a single frequency appears in the Hamiltonian. However, even without rf irradiation, multiple frequencies might be required for example to describe the rotational-resonance (R\textsuperscript{2}) experiment \cite{Colombo:1988fs, Raleigh:1988jh, Levitt:1990ud} where a chemical-shift interaction frame is often used to avoid combined transitions in spin and Fourier space.

In many experiments involving sample rotation and rf irradiation on a single channel, at least two frequencies arise that can be commensurate (rotor-synchronized rf irradiation) or incommensurate.  For cyclic pulse sequences, i.e., a sequence that has no net rotation, a single frequency is sufficient to describe the interaction-frame transformation.
Examples of such experiments that require two frequencies include continuous-wave (cw) irradiation, phase-modulated Lee–Goldburg (PMLG) decoupling \cite{Vinogradov:1999tm, Vinogradov:2000wq}, symmetry-based recoupling sequences \cite{Levitt2007}, or XiX irradiation \cite{Tekely:1994wm, Detken2002} for heteronuclear decoupling.

Experiments with three independent frequencies include MAS with constant irradiation on two different spin species. There are experiments involving MAS with a constant rf irradiation on two different spins, like in Hartmann-Hahn cross-polarization (CP) \cite{Hartmann:1962td, Pines:1973wh, Stejskal:1977wt} or a combination of constant rf irradiation with a cyclic multiple-pulse sequence as in the resonant second-order recoupling (RESORT) experiment \cite{Scholz:2010eg}. Pulse sequences that are not cyclic, i.e., sequences that have a net rotation, always require two frequencies to characterize the interaction frame. Such sequences require three frequencies in combination with MAS, such as two-pulse phase-modulated (TPPM) decoupling \cite{Bennett:1995wr,Scholz:2009bh}. If pulse imperfections like rf-field inhomogeneity, pulse transients or chemical-shift offsets are taken into account in the interaction-frame transformation, even cyclic sequences acquire a net rotation and require two independent frequencies. Experiments involving cw irradiation and a chemical-shift interaction-frame transformation, such as the mixed rotational and rotary-resonance recoupling (MIRROR) sequence \cite{Scholz:2008hr, Scholz:2009dk, Wittmann:2014go, Chavez2022}, require also three frequencies. Additionally, if the influence of the isotropic chemical shift is considered, the number of independent frequencies increases and examples with up to five frequencies can be found in the literature \cite{Shankar:2017ck}.

Various approaches can be employed to handle time-dependent Hamiltonians and derive a propagator for the entire sequence.
The most commonly used method in NMR is average-Hamiltonian theory (AHT) \cite{Haeberlen1976, Mehring:1983wm}, based on the Magnus expansion.
AHT approximates the periodic time-dependent Hamiltonian by a series of time-independent average Hamiltonians over integer multiples of the cycle time.
However, AHT has limitations; it requires defining a single basic frequency and cycle time for the Hamiltonian, and the time evolution of the density operator is accurately described only at multiples of the cycle time.
Consequently, when dealing with multiple incommensurate time-dependent processes in solid-state NMR, such as sample rotation and non-synchronized rf irradiation, AHT is often impractical unless the time scales of the two processes are well-separated and sequential averaging can be applied.

In contrast, Floquet theory \cite{Shirley:1965tn, Vega:2007co, Scholz:2010, Leskes:2010dx, Ivanov:2021uy} offers a more comprehensive approach to describe the time evolution of the spin system at all times and can handle multiple incommensurate frequencies.
Nonetheless, the matrix representation of the full Floquet Hamiltonian is infinite-dimensional and can be challenging to intuitively comprehend its implications on the time evolution of the spin system.
An operator-based formulation of the Floquet Hamiltonian, introduced by Boender \cite{Boender:1996wu} and Augustine \cite{Augustine:1995uf}, presents an opportunity to employ analytical block-diagonalization techniques like the van Vleck transformation \cite{VanVleck:1948we, Primas:1961vi, Primas:1963tg, Goldman:1992td}.
This leads to an approximately diagonalized Floquet Hamiltonian \cite{Vinogradov:2001ui, Ramesh:2001uz, Ernst:2005ic},  which can then be projected back into Hilbert space to obtain an effective time-independent Hilbert-space Hamiltonian.
This effective Hamiltonian allows for a more intuitive discussion of the time evolution of the spin system \cite{Ernst:2005ic}.

Multi-mode operator-based Floquet theory becomes particularly important for dealing with non-resonant problems, which can be challenging to analyze using average-Hamiltonian theory (AHT).
Non-resonant experiments, such as homonuclear \cite{Vinogradov:2004es} or heteronuclear \cite{Madhu:2013df} decoupling using continuous-wave (cw) irradiation or multiple-pulse sequences, fall into this category.
Using Floquet theory to derive effective Hamiltonians, the analysis of experiments involving resonant and non-resonant second-order effects becomes more straightforward.
Such second-order polarization-transfer schemes are less susceptible to dipolar-truncation effects \cite{Hohwy:1999bt, Hohwy:1999ws, Grommek:2006gv, Bayro:2009ci} compared to first-order recoupling sequences and are used to obtain long-range distance restraints in proteins \cite{Manolikas:2008wy, DePaepe:2008bm}, making them valuable for applications in biological solid-state NMR.

However, an important limitation of  Floquet theory lies in its requirement for a periodic Hamiltonian. Consequently, Floquet theory always describes the periodic continuation of the actual Hamiltonian, implying a continuous repetition of the experiment.
In the course of this paper, it will become evident that certain observable effects are not fully captured by traditional Floquet theory due to this mandatory periodic continuation.
Notably, effects such as the width of resonance conditions, recoupling and decoupling performance under MAS detuning, and chemical-shift selective recoupling \cite{Potnuru2021, Xiao2021, Zhang:2019ix, Zhang2020, Chavez2022a}
are often difficult to describe using Floquet theory.
Another shortcoming of Floquet theory (and AHT) is the inability to adequately describe near-resonance conditions due to the slow convergence of the non-resonant series expansion.
However, such descriptions are important for the numerical optimization of effective Hamiltonians.
One possible workaround is the use of the nearest resonance condition and adding the detuning from the resonance condition to the effective Hamiltonian \cite{Shankar:2017ck}.
However, such an approach is only possible for continuous-wave and not for amplitude or phase-modulated irradiation.

Continuous Floquet theory \cite{Chavez2022} offers a more complete description by eliminating the requirement of a periodic continuation of the Hamiltonian.
This framework describes the original Hamiltonian itself, rather than its periodic continuation, introducing the duration of the Hamiltonian as a parameter.
The finite and duration-dependent resonance conditions that arise provide a comprehensive explanation and quantification of various effects.
These include performance deterioration resulting from magic-angle spinning (MAS) or rf-field amplitude mismatch, as well as the impact of chemical-shift offset and incomplete averaging \cite{Chavez2022a}.

Nonetheless, the theory outlined in \cite{Chavez2022} does have a limitation: it amalgamates distinct modulations, such as sample rotation, rf-field irradiation, and chemical shift, into a singular basic frequency, unlike traditional Floquet theory, which explicitly segregates these modulations using distinct basic frequencies.
Additionally, the effective Hamiltonians of varying orders are not readily available in closed forms; their computation entails the evaluation of integrals, rendering their application less convenient.

In the first part of this paper, we tackle these limitations by developing closed-form expressions for the effective Hamiltonians, without the introduction of assumptions or approximations.
Furthermore, we establish an explicit correlation between the various modulations using multiple basic frequencies, akin to the effective Hamiltonians found in the traditional Floquet theory.

In the second part of the paper, we apply the derived effective Hamiltonian to a range of solid-state NMR experiments, spanning from rotary resonance experiments and extending to multi-pulse recoupling sequences such as symmetry-based $\mathrm{C}$- and $\mathrm{R}$-type pulse schemes. 
Within the context of continuous Floquet theory, we investigate how the width of the resonance conditions is influenced by the duration of the pulse schemes, providing a comprehensive understanding of their behavior in practical applications.

%
%

\section{Theory or Effective Hamiltonians in Continuous Floquet Theory\label{sec:theory}}

The first and second-order terms of the effective Hamiltonian in continuous Floquet theory are given as \cite{Chavez2022}
\begin{align}
\label{eq:formal-first-order-effective-hamiltonian}
\bar{\HH}^{(1)} &=  \frac{1}{T} \widehat{\HH}(\Omega=0)\\
\bar{\HH}^{(2)} &= -\frac{1}{2\,T} \, PV
\int 
\frac{[\widehat{\HH}(\Omega)
	,\widehat{\HH}(-\Omega)]}
{\Omega}
  \dd\Omega ~.
\label{eq:formal-second-order-effective-hamiltonian}
\end{align}
The quantity $\widehat{\HH}(\Omega)$ is the Fourier transform of the Hamiltonian defined as
\begin{align}
  \label{eq:fourier-transform-definition-1}
\widehat{\HH}(\Omega) 
  &=  \int\limits_{-\infty}^\infty\HH(t)e^{-i \Omega t}\dd t~,\\
  \HH(t)
  &=  \frac{1}{2\pi} \int\limits_{-\infty}^\infty\widehat{\HH}(\Omega)e^{i \Omega t}\dd\Omega~,
    \label{eq:fourier-transform-definition-2}
\end{align}
where $\HH(t)$ is the time-dependent Hamiltonian of the system under investigation and $T$ is the total duration of this Hamiltonian. Note that the normalization of the Fourier transform has been changed compared to the original publication of continuous Floquet theory \cite{Chavez2022} to obtain a simpler agreement with traditional Floquet theory in the case of periodic continuation of a sequence.
These formal definitions of the effective Hamiltonians are not explicitly dependent on experimental parameters that generate the time dependence of the Hamiltonian. A single frequency characterizes multiple time-dependent processes generated, for example, by the sample rotation 
, interaction-frame transformations to describe the  rf-field irradiation 
or the chemical-shift offset 
\\
The first-order effective Hamiltonian of Eq.~(\ref{eq:formal-first-order-effective-hamiltonian}) is fairly straightforward involving only the evaluation of the Fourier transform of the Hamiltonian at frequency $\Omega = 0$. 
The second-order effective Hamiltonian of Eq.~(\ref{eq:formal-second-order-effective-hamiltonian}), is given as a Cauchy principal value \cite{Kanwal1997}, which has to be evaluated to obtain a useful result. 
This can be limiting, especially if we want to design computationally efficient tools because the principal value has to be solved numerically every time we change the experimental parameters. In addition, solving principal values can be challenging, and in general, require special treatments near the pole of the integral kernel. These treatments often imply higher sampling rates of the numerical integration algorithm and, thus, increases the computational cost significantly. 
A closed-form expression of effective Hamiltonian, therefore, increases interpretability and significantly reduces the computational cost. Hence enabling a more efficient and intuitive approach for the application of continuous Floquet theory without impeding its generality.
\\
The effective Hamiltonians at various orders do not rely on any assumptions or approximations.
As a consequence, the duration of the Hamiltonian $T$ is introduced in the effective theory, which is missing in traditional Floquet theory, because we assume that the Hamiltonian is periodic. In other words, we approximate the 'real' Hamiltonian with a Hamiltonian, which is periodically continued.
\\
In the same way, any Hamiltonian with a finite duration can be expressed as a section of a periodic Hamiltonian. In the following, we are going to explore this idea of periodic continuation and the application of a time-window function in combination with continuous Floquet theory. As a result, we will obtain simpler and more expressive closed-form expressions for the effective Hamiltonians.

%
%

\subsection{Periodic Continuation and Time Window Function}
\label{sec:periodic-continuation}
\begin{figure}[ht]
  \centering
  \includegraphics[width=0.45\textwidth]{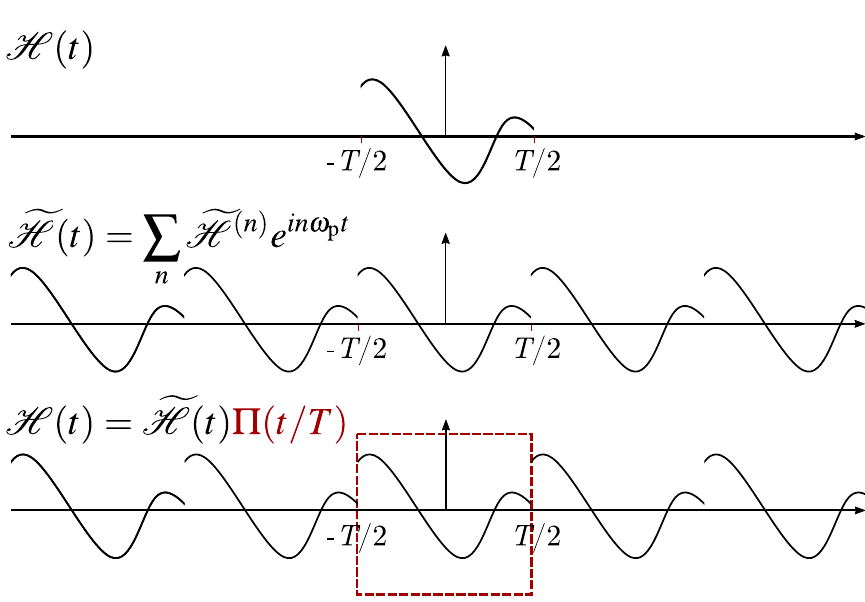}
  \caption{Illustration of the reformulation of a time-dependent Hamiltonian in terms of a periodic continuation of the Hamiltonian $\widetilde{\HH}$ and a rectangular window function.
In this example the Hamiltonian is periodically continued with the period $T$, i.e. with the frequency $\omega_{\mathrm{p}}=2\pi/T$.\label{fig:periodic-continuation}}
\end{figure}

\begin{figure}[ht]
  \centering
  \includegraphics[width=0.45\textwidth]{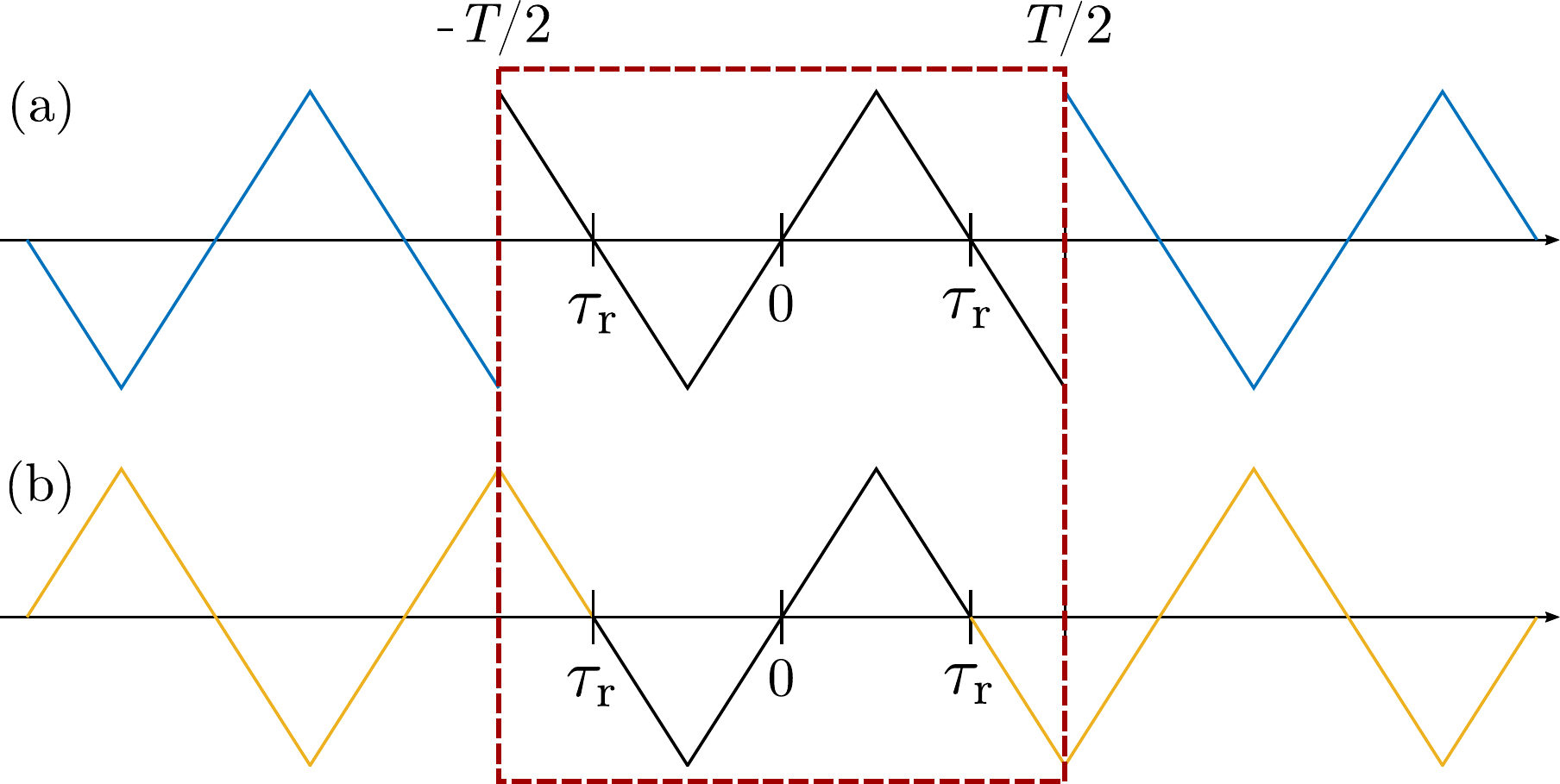}
  \caption{A simplified illustration of two different periodic continuations of the same Hamiltonian. (a) Periodic continuation, where the  Hamiltonian is simply repeated. (b) Periodic continuation with the period $\tau_\mathrm{r}$ of the Hamiltonian that is shorter than the time interval $T$. Both periodic continuations are equal in the relevant interval $t\in[-T/2, T/2]$.  The arising discontinuities in (a) are unproblematic in continuous Floquet theory since we only consider the interval $t\in[-T/2, T/2]$ but will lead to artifacts in traditional Floquet theory since the description outside the relevant interval is not correct. \label{fig:different-periods} }
\end{figure}

Every Hamiltonian of a finite duration can be seen as a section of a Hamiltonian with a longer duration. Thus we can formulate a Hamiltonian as a product of a continuation of itself and a rectangular function representing a time window. At first, this might seem redundant, but allows for substantial simplification of the expressions of the effective Hamiltonians as shown in the following. 
Let us consider a time-dependent Hamiltonian that spans the interval $T$ from $t_0=-T/2$ to $t_{\mathrm{f}} = +T/2$.
In principle, any arbitrary time interval of size $T$ can be chosen, but this choice is symmetric around zero, which will turn out to be convenient later because the Fourier transform of a symmetric function is real. That said, a non-symmetric choice of time interval is possible but will lead to slightly different and, arguably, more complicated expressions for the final effective Hamiltonian. However, the steps of the derivation are in both cases the same and no additional difficulties arise from the choice of symmetry given by the time interval.
As mentioned before, the main idea
is to represent the Hamiltonian as a product of its periodic continuation 
to the entire time domain and a rectangular window function
$\Pi(t/T)$ that 'cuts out' the desired interval $T$: 
\begin{equation}
  \label{eq:general-periodic-continuation}
   \HH(t) = \widetilde{\HH}(t)\,\Pi(t/T)~.
\end{equation}
$\widetilde{\HH}(t)$ denotes a periodic continuation of the Hamiltonian to the entire time domain ($t\in \mathbb{R}$) and $\Pi(t/T)$ is a rectangular window function
\begin{equation}
  \Pi(t/T) =
  \begin{dcases}
    1  & |t| < T/2\\
    0 & \mathrm{else}
  \end{dcases}~.
\end{equation}
Eq.~(\ref{eq:general-periodic-continuation}) is a reformulation of the Hamiltonian and does not introduce any assumption or restrictions on the Hamiltonian.
A graphical illustration of Eq.~(\ref{eq:general-periodic-continuation}) is given in Fig.~\ref{fig:periodic-continuation}.
Note, that in traditional Floquet theory only the periodic continuation $\widetilde{\HH}(t)$ is considered and not the original Hamiltonian.
In addition, we would like to mention, that 
any continuation $\widetilde{\HH}(t)$ satisfies Eq.~(\ref{eq:general-periodic-continuation}), because the window function $\Pi(t/T)$ is only non-zero in the time-interval $t\in [-T/2,T/2]$ of the original Hamiltonian $\HH(t)$.
Hereafter, we consider only periodic continuations of the Hamiltonian, where the Hamiltonian is extended to the entire real domain periodically. A main advantage of using periodic continuations is, that we can formulate it in terms of a Fourier series:
any periodic continued Hamiltonian can be written as a Fourier series with a basic frequency $\omega_{\mathrm{p}} = 2\pi/T$:
\begin{equation}
  \label{eq:Fourier-series}
  \widetilde{\HH}(t) = \sum_n \widetilde{\HH}^{(n)} e^{i  n \omega_{\mathrm{p}} t}
\end{equation}
with
\begin{align}
  \label{eq:fourier-coefficient}
 \widetilde{ \HH}^{(n)}
  &= \frac{1}{T}\int_{-T/2}^{T/2}\HH(t)e^{-i  n \omega_{\mathrm{p}} t}\dd
    t~.
\end{align}
Insertion into Eq.~(\ref{eq:general-periodic-continuation}), immediately leads to the Hamiltonian 
\begin{equation}
  \label{eq:single-frequency-periodic}
  \HH(t) =  \widetilde{\HH}(t)\Pi(t/T)=\sum_n \widetilde{\HH}^{(n)} e^{i n \omega_\mathrm{p} t}\, \Pi(t/T)~.
\end{equation}
Here we considered the simplest periodic continuation, where
the Hamiltonian is extended to the entire time domain by just repeating the original Hamiltonian, depicted schematically in Fig.~\ref{fig:different-periods} a). 
The period of this periodic continuation is the overall duration of the original Hamiltonian $T$, and hence the basic frequency is $\omega_\mathrm{p} = 2 \pi/T$.
Any Hamiltonian can be represented using Eq.~(\ref{eq:single-frequency-periodic}) with the basic frequency $\omega_\mathrm{p}$. While this is simple to implement, it might require high-order Fourier coefficients in $\omega_\mathrm{p}$ because the frequency is not matched to the frequencies of the Hamiltonian and discontinuities are present in the periodic continuation of the Hamiltonian.

If the basic frequencies and the functional form of the Hamiltonian are known, as it is often the case in solid-state NMR, then the periodicity of the Hamiltonian ($\tau_\mathrm{r}$) is known and we can also use a periodic continuation of the Hamiltonian based on the basic frequencies of the Hamiltonian. Fig.~\ref{fig:different-periods} b) shows an example of a different periodic continuation which has the period $\tau_\mathrm{r}<T$. This periodic continuation also reproduces the original Hamiltonian in the relevant interval $t\in [-T/2, T/2]$ and can be used in Eq.~(\ref{eq:single-frequency-periodic}). Such a periodic continuation might be more complex to implement but will require fewer Fourier coefficients to characterize the time-dependent Hamiltonian by avoiding discontinuity. It is important to notice, that a different periodic continuation will always have a different set of Fourier coefficients and Fourier frequencies according to Eq.~(\ref{eq:fourier-coefficient}).

If the Hamiltonian is modulated by several independent processes like MAS and rf irradiation characterized by an interaction-frame transformation, multiple basic frequencies are present and it might be best to define the periodic continuation using a multi-dimensional Fourier series on the set of basic frequencies. 
Further, note that a continuous periodic continuation of $\widetilde \HH$ is not required. Therefore, the discontinuities of the periodic continuation in Fig.~\ref{fig:different-periods} a) do not cause any problems but are rather an inconvenience giving rise to higher-order Fourier coefficients than in the continuous case (Fig.~\ref{fig:different-periods} b)). This could increase the computational cost especially when calculating second- or higher-order terms of the effective Hamiltonian.
This is because the periodic continuation of the Hamiltonian in continuous Floquet theory is only used to reformulate the original Hamiltonian using Eq.~(\ref{eq:single-frequency-periodic}). 
Any periodic extension is applicable; the choice of which one to utilize primarily depends on convenience. In traditional Floquet theory, the situation is different. The periodic continuation of the Hamiltonian serves as an approximation of the original Hamiltonian, i.e., it is the Hamiltonian that we describe in traditional Floquet theory. Therefore we have to ensure that the periodic continuation is a good approximation, which, for example, excludes periodic continuations that are not continuous.
%
%
%

Two primary factors cause the time-dependence of the  Hamiltonians in solid-state NMR: spatial rotation of the sample about one or more axes, and rotations in the spin space caused by rf-field irradiation that can be described by interaction-frame transformations.
Because rotations of the spatial and the spin space are independent of each other, the
Hamiltonian can be formulated as a sum scalar products between the spatial parts  $\vec A_l(t)$ and the spin parts $\vec{\mathcal{I}}_l(t)$, i.e. 
\begin{equation}
  \label{eq:general-Hamiltonian}
  \HH(t) = \sum_{l} \vec A_l(t)\vec{\mathcal{I}}_l(t)~.
\end{equation}
Typically in solid-state NMR the spatial part is modulated due to MAS and can be written as a Fourier series in the MAS frequency $\omega_\mathrm{r}$:
\begin{equation}
  \vec A_l(t) = \sum_{n=-l}^{l} \vec A_l^{(n)} e^{i n \omega_\rr t} ~,
\end{equation}
with
\begin{equation}
  \vec{A}_l^{(n)} = \frac{1}{\tau_\rr}\int_{-\tau_\rr/2}^{\tau_\rr/2} \vec A_l(t) e^{-i n \omega_\rr t} \dd t~.
\end{equation}
In the usual rotating frame, the time dependence of the spin part $\vec{\mathcal{I}}_l(t)$ is often due to the rf-field irradiation which we can describe by an interaction-frame transformation.
For a general periodic sequence that is cyclic, i.e., the interaction frame can be described by an identity propagator over a whole cycle, only a single frequency, $\omega_\mathrm{m}=2\pi/\tau_\mathrm{m}$ is required to characterize the time-dependent interaction-frame Hamiltonian. If the interaction frame is not cyclic, i.e., if the interaction frame has an effective field nutation over the cycle time $\tau_\mathrm{m}$, employing a Fourier series with two frequencies instead of a single basic frequency can be advantageous. This is particularly relevant for multiple-pulse sequences, where the Hamiltonian exhibits multiple time dependencies, and also when dealing with multiple chemical shifts. In these scenarios, the use of multiple frequencies in the Fourier series proves to be more insightful for an accurate periodic continuation of Hamiltonian.
A common choice of frequencies is using the modulation frequency $\omega_{\text{m}}$ of the multi-pulse sequence and the effective field frequency $\omega_\eff=\beta/\tau_{\text{m}}
$, where $\beta$ is the effective flip angle of the periodic sequence.
\begin{equation}
  \label{eq:spin-tensor-two-frequency}
\vec{\mathcal{I}}_{l}(t) = \sum_{k}\sum_{\ell=-l}^{l} \vec{\mathcal{I}}^{(k,\ell)}_{l} e^{i k \omega_\mathrm{m} t}e^{i \ell \omega_\eff t} \Pi(t/T)~,
\end{equation}
with
\begin{equation}
  \label{eq:bimodal-spin-part}
  \vec{\mathcal{I}}^{(k,\ell)}_{l} = \frac{1}{T}\int_{-T/2}^{T/2}\vec{\mathcal{I}}_{l}(t)e^{-i (k \omega_\mathrm{m}+\ell\omega_\eff )t}\dd t~.
\end{equation}
Therefore the complete Hamiltonian can be written as
\begin{align}
   \label{eq:spin-tensor-multiple-frequencies}
  \widetilde{\HH}(t) &= \sum_{l} \sum_{n,k,\ell} \vec A_l^{(n)} e^{i n \omega_\rr t} \vec{\mathcal{I}}^{(k,\ell)}_{l} e^{i (k \omega_\text{m}+ \ell\omega_\eff )t} \notag \\
         &=  \sum_{n,k,\ell} \widetilde{\HH}^{(n,k,\ell)}   e^{i (n \omega_\rr + k \omega_\mathrm{m}+\ell\omega_\eff ) t}
\end{align}
The choice of characteristic frequencies to express the time-dependent Hamiltonian depends on the problem and the interaction frame chosen to describe the problem. For example, only $\omega_{\rr}$ and $\omega_\text{m}$ are required, if we have a cyclic interaction frame, that does not generate an effective field $\omega_{\eff}$. For experiments using different irradiation schemes on multiple spins, multiple interaction frames might expose additional frequencies that modulate the spin part of the Hamiltonian.
For other sample rotation techniques like double rotation, multiple spinning frequencies might be required to describe the spatial part of the Hamiltonian.
Alternatively, it is always possible to use only a single characteristic frequency ($\omega_\mathrm{p} = 2\pi/T$) or the definition of the effective Hamiltonians (Eqs.~(\ref{eq:formal-first-order-effective-hamiltonian}) and (\ref{eq:formal-second-order-effective-hamiltonian})) directly.

Multi-index notation provides a concise way to write the Hamiltonians, independent of the chosen periodic continuation and choice of characteristic frequencies:
\begin{equation}
\label{eq:spin-tensor-general-frequencies}
  \HH(t)  =  \widetilde{\HH}(t) \Pi(t/T) =  \sum_{\mathbf{n}} \widetilde{\HH}^{(\mathbf{n})}   e^{i \omega_\mathbf{n} t} \Pi(t/T)~.
\end{equation}
Here $\mathbf{n}$ is the multi index, which is a tuple containing all the indices.
For Eq.~(\ref{eq:spin-tensor-multiple-frequencies}) the multi index is given as $\mathbf{n} = (n,k,\ell)$ and $\omega_{\mathbf{n}} = n \omega_\mathrm{r} + k \omega_\mathrm{m}+ \ell \omega_\eff$.
It is important to recognize, that $\omega_{\mathbf{n_0}} = 0$ are the resonance conditions, where $\mathbf{n}_0$ indicate the multi indices that lead to a resonance condition. 
\\
The appearance of specific characteristic frequencies in the effective Hamiltonian is a matter of choice and different choices are always possible. This is similar to the choice of coordinate system or interaction frame. Although many choices are possible, only a few are convenient for a specific problem. Usually, the choice of a suitable set of characteristic frequencies is straightforward and arises naturally as will become evident in Section \ref{sec:application}.

%
%

\subsection{Effective Hamiltonian}
Based on the general representation of the time-dependent Hamiltonian given in Eq.~(\ref{eq:spin-tensor-general-frequencies}) we can simplify the effective Hamiltonians (Eqs.~(\ref{eq:formal-first-order-effective-hamiltonian}) and (\ref{eq:formal-second-order-effective-hamiltonian})) significantly. The resulting effective Hamiltonians are simple closed-form expressions that are generally applicable.
Before we can begin to derive the actual effective Hamiltonian we have to calculate the Fourier transformation of the time-dependent Hamiltonian to obtain the frequency-space Hamiltonian $\widehat{\HH}(\Omega)$.
For the (element-wise) Fourier transform (Eq.~(\ref{eq:fourier-transform-definition-1})) of the Hamiltonian (Eq.~(\ref{eq:spin-tensor-general-frequencies})), it follows immediately
\begin{equation}
  \label{eq:frequency-domain-Hamiltonian}
  \widehat{\HH}(\Omega)
  = T \sum_{\mathbf{n}} \HH^{(\mathbf{n})} \mathrm{sinc}\left(\frac{(\omega_\mathbf{n}-\Omega )T}{2}\right) 
\end{equation}
where $\mathrm{sinc}\left(x\right)=\sin(x)/x$. For the sake of clarity, the derivation is not shown here but can be found in the SI.
Note, that we have an explicit dependence on $T$, $\omega_{\mathbf{n}}$ and $\Omega$ in the factor $\mathrm{sinc}\left((\omega_{\mathbf{n}}-\Omega)T/2\right)$. The representation of the Hamiltonian in Eq.~(\ref{eq:frequency-domain-Hamiltonian}) is independent of the details of the frequencies required to describe the Hamiltonian.
This is important because it allows us to obtain general closed-form solutions of the effective Hamiltonians since we can carry out the integration for $\Omega$ in the definition of the effective Hamiltonians (Eq.~(\ref{eq:formal-first-order-effective-hamiltonian}) and (\ref{eq:formal-second-order-effective-hamiltonian})), as shown in the following two sections.

\subsubsection{First-order effective Hamiltonian}
\begin{figure*}[ht]
  \centering
  \includegraphics[width=0.9\textwidth]{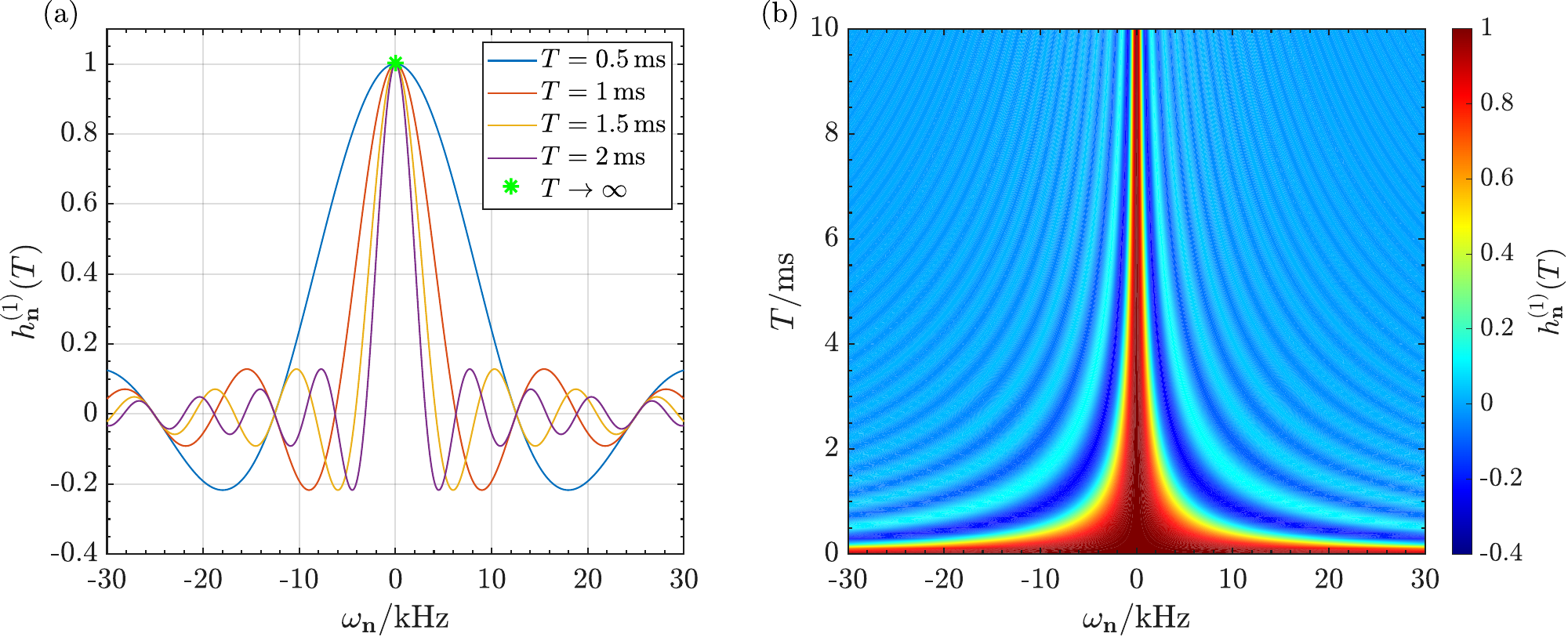}
  \caption{Plots of the function $h^{(1)}_{\mathbf{n}}(T)$. (a) $h^{(1)}_{\mathbf{n}}(T)$ as function of the resonance $\omega_{\mathbf{n}}$ for different $T$. For $T\rightarrow\infty$ we obtain the result known from traditional Floquet theory, depicted as a green star. (b) $h^{(1)}_{\mathbf{n}}(T)$ as function of $\omega_{\mathbf{n}}$ and $T$. The longer the duration $T$ becomes, the narrower the function $h^{(1)}_{\mathbf{n}}(T)$, i.e., the more localized is the contribution of the first-order Hamiltonian. \label{fig:h1}}
\end{figure*}

The first-order effective Hamiltonian can simply be obtained by 
inserting the Fourier Series of (Eq.~(\ref{eq:frequency-domain-Hamiltonian})) in the formal definition of the first-order effective Hamiltonian Eq.~(\ref{eq:formal-first-order-effective-hamiltonian}), leading to
\begin{align}
  \label{eq:first-order-effective-hamiltonian}
  \bar{\HH}^{(1)}= \frac{1}{T} \widehat{\HH}(0)
  &= \sum_{\mathbf{n}} \widetilde \HH^{(\mathbf{n})}\,h^{(1)}_{\mathbf{n}}(T)~,
\end{align}
with
\begin{equation}
  \label{eq:h1}
  h^{(1)}_{\mathbf{n}}(T) = \text{sinc}
  \left(\frac{ \omega_{\mathbf{n}} T }{2}\right)~. 
\end{equation}
Eq.~(\ref{eq:first-order-effective-hamiltonian}) is an expansion of the first-order effective Hamiltonian in terms of sinc functions $h^{(1)}_{\mathbf{n}}(T)$, similar to the Whittaker–Shannon interpolation formula \cite{Whittacker1915, Shannon1998}.
Fig.~\ref{fig:h1} shows the function $h^{(1)}_{\mathbf{n}}(T)$ as function of $\omega_\mathbf{n}$ for different overall duration $T$. The longer the duration $T$ the narrower the function $h^{(1)}_{\mathbf{n}}$ becomes.
Note, that in continuous Floquet theory, all Fourier coefficients $\widetilde \HH^{(\mathbf{n})}$ contribute to the first-order effective Hamiltonian for every value of $\omega_{\mathbf{n}}$. Resonant terms ($\omega_{\mathbf{n}_0} =0$) contribute strongly ($\text{sinc}(0)=1$) while non-resonant terms ($\omega_{\mathbf{n}} \neq 0$) contribute weakly with their weight given by the value of the sinc function at frequency zero.
In contrast, the first-order Hamiltonian in traditional Floquet theory only contains resonant terms $\widetilde \HH^{(\mathbf{n}_0)}$ ($\omega_{\mathbf{n}_0} = 0$) and, therefore, does not describe near-resonant behavior as we will see later in various examples.\\

By taking the limit $T\rightarrow\infty$ of Eq.~(\ref{eq:first-order-effective-hamiltonian}) and  Eq.~(\ref{eq:h1}), we obtain:
\begin{equation}
  \label{eq:limit-h1}
  \lim\limits_{T\to \infty} h^{(1)}_{\mathbf{n}}(T) =
  \begin{dcases}
   1 &  \text{if}\quad \omega_\mathbf{n} = \omega_\mathbf{n_0} = 0\\
   0 & \text{else}
  \end{dcases}~,
\end{equation}
which leads to the first-order effective Hamiltonian of single- and multi-mode traditional Floquet theory:
\begin{equation}
  \label{eq:first-order-effective-hamiltonian-formal}
\lim\limits_{T\to \infty} \bar{\HH}^{(1)} = \sum_{\mathbf{n_0}} \widetilde \HH^{(\mathbf{n_0})}~,
\end{equation}
For $T\rightarrow\infty$ the function $h^{(1)}_{\mathbf{n}}(T)$ is only non-zero at the resonance conditions $\omega_\mathbf{n_0}=0$. Therefore the sum of in Eq.~(\ref{eq:first-order-effective-hamiltonian-formal}) is only over multi-index $\mathbf{n_0}$, which fulfills the resonance condition $\omega_{\mathbf{n_0}}=0$. As mentioned before,
the first-order effective Hamiltonians of traditional Floquet theory, only leads to correct predictions at the resonance conditions $\omega_{\mathbf{n_0}}=0$, but not for $\omega_{\mathbf{n_0}}\neq0$. In the following, we show that this is due to the requirement of periodic Hamiltonians in traditional Floquet theory. 
\subsubsection{Second-order effective Hamiltonian}
\begin{figure*}[ht]
  \centering
  \includegraphics[width=0.9\textwidth]{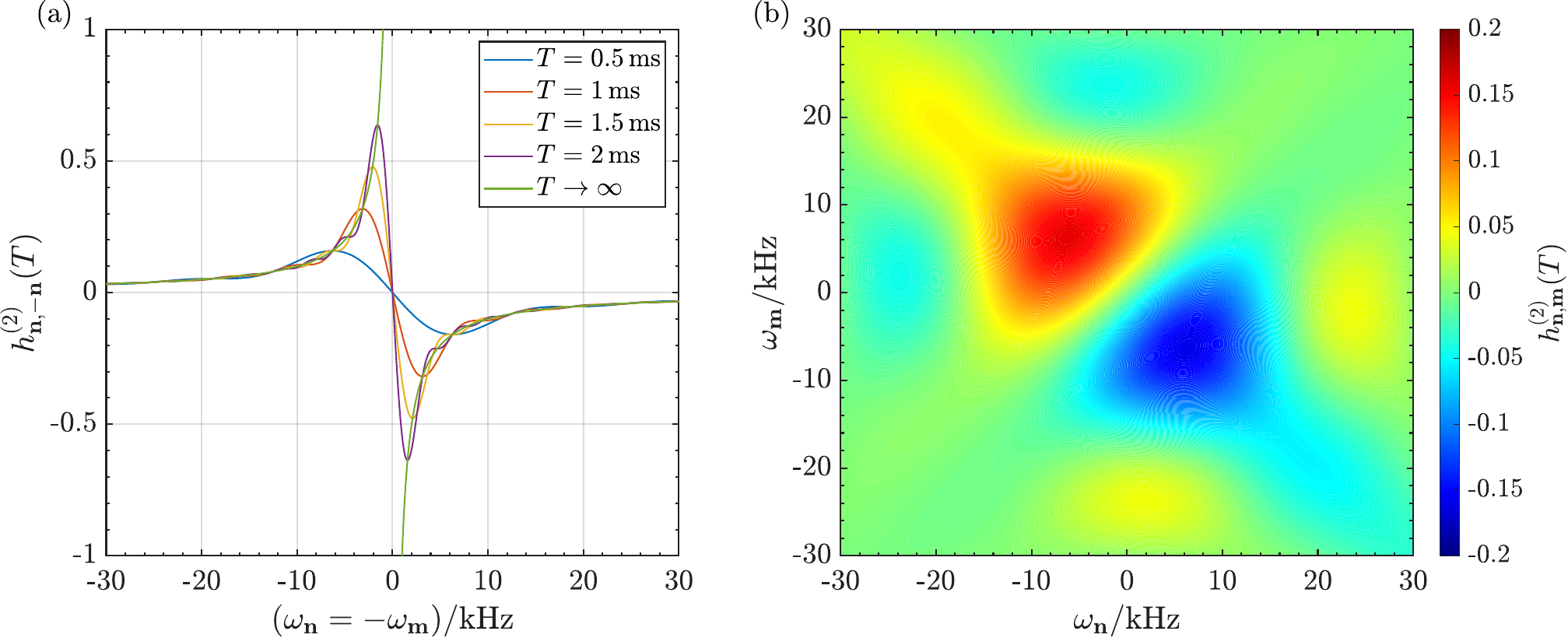}
  \caption{Plots of the function $h^{(2)}_{\mathbf{n},\mathbf{m}}(T)$. (a) $h^{(2)}_{\mathbf{n},-\mathbf{n}}(T)$ as function of the resonance $\omega_{\mathbf{n}}=\omega_{-\mathbf{m}}$ for different $T$. Note, that for $T\rightarrow\infty$ we obtain the result known from traditional Floquet theory (green line). (b) $h^{(2)}_{\mathbf{n},\mathbf{m}}(T)$ as function of $\omega_{\mathbf{n}}$ and $\omega_{\mathbf{m}}$ for $T=0.5\,$ms. Note, that the function is anti-symmetric in respect to the $\omega_{\mathbf{n}} = \omega_{\mathbf{m}}$ axis.}
  \label{fig:h2}
\end{figure*}

Let us now turn to the derivation of the closed-form expression of the second-order Hamiltonian.
Again, we use Eq.~(\ref{eq:frequency-domain-Hamiltonian}) to evaluate the integral in the definition of the second-order effective Hamiltonian Eq.~(\ref{eq:formal-second-order-effective-hamiltonian}).
For the second-order effective Hamiltonian follows
\begin{align}
  \label{eq:second-order-effective-hamiltonian}
    \bar{\HH}^{(2)}
  &= -\frac{1}{2}\sum_{\mathbf{n},\mathbf{m}}[\widetilde{\HH}^{(\mathbf{n})},\widetilde{\HH}^{(\mathbf{m})}]
    \,h^{(2)}_{\mathbf{n},\mathbf{m}}(T)~,
\end{align}
with
\begin{widetext}
\begin{equation}
  \label{eq:basis-function2}
  h^{(2)}_{\mathbf{n},\mathbf{m}}(T) = \frac{2}{T} \frac{ \omega_\mathbf{m} \sin \left(\frac{\omega_\mathbf{n} T}{2}\right) \cos \left(\frac{\omega_\mathbf{m} T}{2}\right)-\omega_\mathbf{n} \cos \left(\frac{\omega_\mathbf{n} T}{2}\right) \sin \left(\frac{\omega_\mathbf{m} T}{2}\right)}
  {\omega_\mathbf{n} \omega_\mathbf{m} (\omega_\mathbf{n}+\omega_\mathbf{m})}
  ~.
\end{equation}
\end{widetext}
The derivation of these equations can be found in the SI. 
Note, that the singularities of the functions $h^{(2)}_{\mathbf{n},\mathbf{m}}(T)$ in $\omega_\mathbf{n}$ and $\omega_\mathbf{m}$ are removable, i.e.,
\begin{align}
  &\lim_{\omega_{\mathbf{n}/\mathbf{m}} \to 0 }h^{(2)}_{\mathbf{n},\mathbf{m}}(T) = \pm \frac{
    \omega_{\mathbf{n}/\mathbf{m}} T \cos \left(\frac{\omega_{\mathbf{n}/\mathbf{m}} T}{2}\right)-2 \sin \left(\frac{\omega_{\mathbf{n}/\mathbf{m}} T}{2}\right)}{\omega_{\mathbf{n}/\mathbf{m}}^2 }~,
\\
  &\lim_{\omega_{\mathbf{n}/\mathbf{m}} \to -\omega_{\mathbf{m}/\mathbf{n}}} h^{(2)}_{\mathbf{n},\mathbf{m}}(T)= 
    \pm  \frac{  \sin (\omega_{\mathbf{m}/\mathbf{n}}T)-\omega_{\mathbf{m}/\mathbf{n}} T}{\omega_{\mathbf{m}/\mathbf{n}}^2 }~.
\end{align}
Figure~\ref{fig:h2} (a) shows $h^{(2)}_{\mathbf{n},-\mathbf{n}}(T)$ as a function of $\omega_{\mathbf{n}}=-\omega_{\mathbf{m}}$ for different durations $T$. For increasing duration the $h^{(2)}_{\mathbf{n},\mathbf{m}}(T)$ approximates $\omega^{-1}_\mathbf{n}$. In contrast, for $T\rightarrow 0$, $h^{(2)}_{\mathbf{n},\mathbf{m}}(T)$ approaches $0$ as required for a second-order term.
Similar to the first-order effective Hamiltonian, the second-order effective Hamiltonian leads to a description of both resonant ($\omega_{\mathbf{n_0}}=0$) and non-resonant ($\omega_{\mathbf{n}}\neq0$) behavior. 
Again we can show the correspondence to traditional Floquet theory by taking the limit $T\rightarrow\infty$:
\begin{equation}
  \label{eq:limit-h1}
  \lim\limits_{T\to \infty} h^{(2)}_{\mathbf{n},\mathbf{m}}(T) =
  \begin{dcases}
   {\omega_\mathbf{n}^{-1}} & \text{if} \quad \omega_\mathbf{n} = -\omega_\mathbf{m}\\
   0 & \text{else}
  \end{dcases}~,
\end{equation}
Hence we obtain
\begin{align}
      \bar{\HH}^{(2)}
  & = -\frac{1}{2}\sum_{\mathbf{n_0}}\sum_{\mathbf{n}\neq \mathbf{n_0}}
    \frac{[\HH^{(\mathbf{n})},\HH^{(\mathbf{n_0-n})}]}{\omega_{\mathbf{n}}}
\end{align}
This is exactly the second-order effective Hamiltonian as obtained by traditional operator-based Floquet theory, written with multi-index notation.
Note, that in the limit process the two multi-indices $\mathbf{n}, \mathbf{m}$ present in Eq.~(\ref{eq:second-order-effective-hamiltonian}) are reduced to only one, i.e., $\mathbf{n}$, for traditional Floquet theory and that the resonance condition $\omega_{\mathbf{n_0}}=0$ is excluded in the summation. However, in continuous Floquet theory, $\omega_{\mathbf{n_0}}=0$ is not excluded but has zero contribution due to the zero value of $h^{(2)}_{\mathbf{n},\mathbf{-n}}(T)$.

As mentioned a particular effective Hamiltonian corresponds to a specific duration $T$.
To change the duration of an effective Hamiltonian, we just vary the parameter $T$, which only affects the functions $h_{\mathbf{n}}^{(1)}$ and $h^{(2)}_{\mathbf{n,m}}$ of the first- and second-order effective Hamiltonian. Hence the effective Hamiltonians of continuous Floquet theory are \textit{not} linear in $T$.

%
%
\section{Application\label{sec:application}}
The following examples serve two purposes: firstly, as a practical guide on how to apply continuous Floquet theory to various experimental schemes, and secondly, to demonstrate the importance of the finite duration of pulse sequences for the description of near-resonance behavior.
The procedure for applying continuous Floquet theory consists of four main steps:
\begin{enumerate}
\item[(i)]
Define the extent of your spin system that is required for the experiment to be described.
One strategy is to start with a minimal configuration and gradually adding complexity to study its effects. This approach offers flexibility and reduces computational burden. Alternatively, one can begin with a more general model and simplify iteratively. While more effort may be required initially, it ensures a thorough understanding and adaptability to different experimental conditions.
\item[(ii)] Identify and apply a suitable interaction-frame transformation \cite{Cho1999, Boutis:2003bi, Chavez2022a}. This step allows us to adapt the theory to the specific problem at hand by selecting an interaction frame that is most suitable for observing the dynamics of interest and simplifying the Hamiltonian. In many cases, this is the rf irradiation and can in addition include the chemical shift. 
\item[(iii)] Choose a set of characteristic frequencies to formulate a periodic continuation of the Hamiltonian as a Fourier series. In this step, there is flexibility to customize the description according to the system of interest by selecting appropriate characteristic frequencies for the Fourier series expansion of the Hamiltonian. As a default option, continuation with the frequency $\omega_\mathrm{p}=2\pi/T$ (Eq.~(\ref{eq:single-frequency-periodic})) can always be used.
\item[(iv)] Insert the Fourier coefficients of the Hamiltonian into the expressions of the effective Hamiltonians. This step involves substituting the Fourier coefficients of the Hamiltonian into the definition of the effective Hamiltonians (Eqs.~(\ref{eq:first-order-effective-hamiltonian}) and~(\ref{eq:second-order-effective-hamiltonian})).
\end{enumerate}
The general Hamiltonian in solid-state NMR consists of a $\II_{N_I}\SSS_{N_S}$ system of spin-1/2 which can be described by a time-dependent Hamiltonian of the form:
\begin{align}
  \label{eq:heteronuclear-spin-system}
  \HH(t) &= \HH_{\II\II}(t) + \HH_{\II\SSS}(t) + \HH_{\SSS\SSS}(t)\\
  &\quad+ \HH_{\II}(t) + \HH_{\SSS}(t) + \HH_\mathrm{rf}(t)~.
\end{align}
The chemical shifts are characterized with
\begin{align}
  \HH_{\II}(t) &= \sum\limits_{p}\sum\limits_{n = -2}^2  \omega_{\II_p}^{(n)}e^{i n \omega_{\rr} t}\II_{pz}~,\\
  \HH_{\SSS}(t) &= \sum\limits_{p}\sum\limits_{n = -2}^2  \omega_{\SSS_p}^{(n)}e^{i n \omega_{\rr} t}\SSS_{pz}~,
\end{align}
where $n=0$ describes to the isotropic and $n\neq0$ the anisotropic chemical shift.
The homonuclear couplings between the $\II$ and $\SSS$ spins are given by
\begin{align}
  \HH_{\II\II}(t) &=\sum_{p<q}\biggl[\omega_{\II_p \II_q}^{(0)}\,2\,\vec{\II}_p\cdot \vec{\II}_q \notag\\
  &\quad\quad\quad + \sum\limits_{\substack{n = -2\\n\neq0}}^2 \omega_{\II_p \II_q}^{(n)}e^{i n \omega_{\rr} t}\,(3 \, \II_{pz}\II_{qz}-\vec{\II}_p\cdot \vec{\II}_q)\biggr] ~,\\
  \HH_{\SSS\SSS}(t) &= \sum_{u<v}\biggl[\omega_{\SSS_u \SSS_v}^{(0)}\,2\,\vec{\SSS}_u\cdot \vec{\SSS}_v \notag\\
  &\quad\quad\quad + \sum\limits_{\substack{n = -2\\n\neq0}}^2 \omega_{\SSS_u \SSS_v}^{(n)} e^{i n \omega_{\rr} t}\,(3 \, \SSS_{uz}\SSS_{vz}-\vec{\SSS}_u \cdot \vec{\SSS}_v)\biggr] ~.
\end{align}
The $\omega_{\II_p\II_q}^{(0)}$ corresponds to the isotropic $J$ coupling and $\omega_{\II_p \II_q}^{(n)}$ to the anisotropic dipolar coupling.
\begin{align}
  \HH_{\II\SSS}(t) &= \sum\limits_{p,u}\sum\limits_{n = -2}^2 \omega_{\II_p\SSS_u}^{(n)}e^{i n \omega_{\rr} t}\,2 \, \II_{pz}\SSS_{uz}
\end{align}
characterizes the isotropic ($n=0$) and an isotropic ($n\neq0$) heteronuclear coupling. Expressions for the various $\omega^{(n)}$ parameters can be found in Ref. \cite{Scholz:2010} Eqs. (34) and (35). Finally,
\begin{align}
  \HH_\mathrm{rf}(t) &= \omega_{1\II}(t) \sum\limits_{p} (\II_{px} \cos(\phi_\II(t))+\II_{py} \sin(\phi_{\II}(t))) \notag\\
  &\quad +\omega_{1\SSS}(t) \sum\limits_{u} (\SSS_{ux} \cos(\phi_\SSS(t))+\SSS_{uy} \sin(\phi_\SSS(t)))
\end{align}
describes arbitrary rf irradiation on the $\II$ and $\SSS$ spins, respectively.
%
%
\subsection{Recoupling Using Constant RF Irradiation Under MAS}
NMR experiments often involve applying a constant rf irradiation during a time $T$ to a single spin species, such as in recoupling experiments like homonuclear rotary resonance (HORROR) \cite{Nielsen:1994tz} or rotary-resonance recoupling ($\mathrm{R}^3$) \cite{Levitt:1988uk,Oas:1988vc}, as well as non-resonant continuous wave (CW) decoupling experiments \cite{Ernst:2005ic}. 
In the following, we will consider the Hamiltonian of Eq.~(\ref{eq:heteronuclear-spin-system}) with a constant rf irradiation on the $\II$ spin with duration $T$, i.e.,
\begin{equation}
  \HH_{\rf}(t) = \omega_1 \II_x\, \Pi(t/T)~.
\end{equation}
As a first step, we simplify the Hamiltonian of Eq.~(\ref{eq:heteronuclear-spin-system}) by going into an interaction frame with the rf irradiation mediated by the unitary operator
\begin{equation}
  \UU(t) = \exp\left(i \omega_1 t \sum_p \II_{pz} \right)
  \exp\left(-i \frac{\pi}{2} \sum_p \II_{py} \right)~.
\end{equation}
This operator first rotates the Hamiltonian by
90$^\circ$ about the $-y$ axis $(\II_{p x} \rightarrow  \II_{p z}, \II_{p z} \rightarrow  - \II_{p x})$ and subsequently transforms the Hamiltonian with the radio-frequency part of the Hamiltonian into an interaction frame. Note, that alternatively the chemical-shift term can also be included in the interaction-frame transformation. In this case, the chemical-shift term is eliminated but manifests as a change in the modulation of the coupling terms. Hence, on one hand, the Hamiltonian becomes simpler because only the interaction-frame modulated coupling terms are left, but on the other hand, it adds complexity to the time dependence of the coupling terms. Here, the chemical shift is not included to prevent obfuscation of the simple modulation of the coupling terms caused by the rf irradiation. Note that for complex rf irradiations or to study the chemical-shift dependence, it is often advisable to eliminate the chemical-shift term by including it in the interaction frame.
After the interaction-frame transformation, we can identify two intrinsic frequencies, namely $\omega_\rr$ and $\omega_{1}$, which naturally serve as suitable options for the Fourier series expansion of the  Hamiltonian. In principle, all terms in the interaction frame should be marked by a $'$ to distinguish them from the standard rotating frame but for ease of notation, we drop the $'$ again after Eq.~(\ref{eq:CW}).
\begin{align}
  \label{eq:CW}
  \HH'(t)&=\UU(t) \HH(t) \UU^\dagger(t)  \notag\\
  &= \sum_{n=-2}^2 \sum_{k=-2}^2 \HH'^{(n, k)} \, e^{i n \omega_{\mathrm{r}} t}\,  e^{i k \omega_1 t}\, \Pi(t/T)~,
\end{align}
with the Fourier coefficients
\begin{align}
  {\HH}^{(0,0)}&= \sum_{u} \omega_{\SSS_{u}}^{(0)}\SSS_{uz}
                         +\sum_{p<q}  \omega_{\II_p \II_q}^{(0)} 2 \vec{\II}_{p} \cdot \vec{\II}_q\notag\\
  &\quad +\sum_{u<v}  \omega_{\SSS_{u} \SSS_{v}}^{(0)} 2 \vec{\SSS}_{u} \cdot \vec{\SSS}_{v}~,\label{eq:general-fourier-00}\\[0.1cm]
  {\HH}^{(n, 0)}&= \sum_{u} \omega_{S_{u}}^{(n)}\SSS_{uz}
                          -\frac{1}{2}\sum_{p<q} \omega_{\II_{p} \II_q}^{(n)}\left[3{\II}_{p z} {\II}_{q z}
                          -\vec{\II}_p \cdot \vec{\II}_q\right]\notag\\
                          &\quad+\sum_{u<v} \omega_{\SSS_{u}\SSS_v}^{(n)}\biggl[3{\SSS}_{u z} {\SSS}_{v z}
                          -\vec{\SSS}_u \cdot \vec{\SSS}_v\biggr]~,\label{eq:general-fourier-n0}\\[0.1cm]
  {\HH}^{(n, \pm 1)}&=-\sum_{p} \frac{\omega_{\II_{p}}^{(n)}}{2} {\II}_{p}^{\pm}
                           - \sum_{p u}\omega^{(n)}_{\II_p \SSS_u} \II_{pz}\SSS^{\pm}_{u}~,\label{eq:general-fourier-n1}\\[0.1cm]
  {\HH}^{(n, \pm 2)}&= -\sum_{p<q} \frac{3}{4}  \omega_{\II_{p} \II_{q}}^{(n)}{\II}^{\pm}_{p} {\II}^{\pm}_{q}~.\label{eq:general-fourier-n2}
\end{align}
From Eq.~(\ref{eq:first-order-effective-hamiltonian}) we directly obtain the first-order Hamiltonian
\begin{equation}
   \label{eq:H1-rotary}
  \bar{\HH}^{(1)} = \sum_{n,k} \HH^{(n,k)} h^{(1)}_{(n,k)}(T)~,
\end{equation}
with
\begin{equation}
  \label{eq:h1-rotary}
  h^{(1)}_{(n,k)}(T) = \text{sinc}\left(\frac{(n\omega_\rr+k\omega_1)T}{2}\right)~.
\end{equation}
The second-order effective Hamiltonian follows directly from Eq.~(\ref{eq:second-order-effective-hamiltonian})
\begin{align}
    \label{eq:H2-rotary}
   \bar{\HH}^{(2)} = -\frac{1}{2}\sum_{\substack{n_1,k_1\\n_2,k_2}} [\HH^{(n_1,k_1)},\HH^{(n_2,k_2)}] h^{(2)}_{(n_1,k_1),(n_2,k_2)}(T)~,
 \end{align}
 with
\begin{widetext}
\begin{equation}
   \label{eq:h2-rotary}
   h^{(2)}_{(n_1,k_1),(n_2,k_2)}(T) = \frac{2}{T} \frac{ \omega_{n_2,k_2} \sin \left(\frac{\omega_{n_1,k_1} T}{2}\right) \cos \left(\frac{\omega_{n_2,k_2} T}{2}\right)-\omega_{n_1,k_1} \cos \left(\frac{\omega_{n_1,k_1} T}{2}\right) \sin \left(\frac{\omega_{n_2,k_2} T}{2}\right)}
   {\omega_{n_1,k_1} \omega_{n_2,k_2} (\omega_{n_1,k_1}+\omega_{n_2,k_2})}
\end{equation}
\end{widetext}
where $\omega_{n_1,k_1} = (n_1\omega_\rr+k_1\omega_1)$ and $\omega_{n_2,k_2} = (n_2\omega_\rr+k_2\omega_1)$.
In classical Floquet theory, only the resonant terms fulfilling the conditions $n_0\omega_\mathrm{r}+k_0\omega_1=0$ are included in the first-order effective Hamiltonian \cite{Ernst:2005ic}, since all other terms do not contribute. For continuous Floquet theory, all Fourier coefficients will contribute to the first-order effective Hamiltonian with the magnitude determined by the width of the sinc function and the distance of the frequency $n\omega_\mathrm{r}+k\omega_1$ from the frequency 0.
Therefore, for fast MAS, the terms $\widetilde{{\HH}}^{(n, 0)}$ never contribute significantly to the first-order effective Hamiltonian while the term $\widetilde{{\HH}}^{(0,0)}$ always contributes but is typically small and can often be neglected.
In the following, we will consider the recoupling effects governed by the resonance conditions:
\begin{enumerate}
\item[(i)] At $\omega_1 = \omega_\rr/2$ only the homonuclear dipolar coupling is recoupled. This condition is called the HORROR condition \cite{Nielsen:1994tz}, and the most significant Fourier coefficients here are  $\HH^{(0,0)},\HH^{(1,-2)}$ and $\HH^{(-1,2)}$.
\item[(ii)] At $\omega_1 = \omega_\rr$ the CSA tensor, the heteronuclear dipolar and the homonuclear dipolar coupling are recoupled \cite{Levitt:1988uk,Oas:1988vc}. Hence the relevant Fourier coefficients are  $\HH^{(0,0)}$, $\HH^{(1,-1)}$, $\HH^{(-1,1)}$, $\HH^{(2,-2)}$ and $\HH^{(-2,2)}$.
\item[(iii)] At $\omega_1 = 2 \omega_\rr$ the CSA tensor and the heteronuclear dipolar coupling \cite{Levitt:1988uk,Oas:1988vc} are recoupled.
Here the Fourier coefficients $\HH^{(0,0)}$, $\HH^{(2,-1)}$ and $\HH^{(-2,1)}$ dominate.
\end{enumerate}
Condition (ii) and (iii) are rotary-resonance conditions, which include all conditions of the form $\omega_1 = n \omega_\rr$ with $n \in \mathbb{N}$. The higher-order rotary-resonance conditions ($n \geq 3$), however, are less relevant because they are much weaker and difficult to observe experimentally. Furthermore, also fractional rotary resonance condition exist at $\omega_1 = k\omega_\rr/n$ which are usually also relatively week \cite{vanBeek:2011dk}.
Because the Fourier coefficient $\HH^{(0,0)}$ is typically weak, $\HH^{(n,\pm 1)}$ and $\HH^{(n,\pm 2)}$ characterize the time evolution caused by the irradiation. The first one represents the recoupling of the CSA and heteronuclear dipolar coupling, while the second one includes the DQ recoupling of the homonuclear dipolar coupling.
%
%
\subsubsection{HORROR}
\begin{figure*}[ht]
  \centering
  \includegraphics[width=\textwidth]{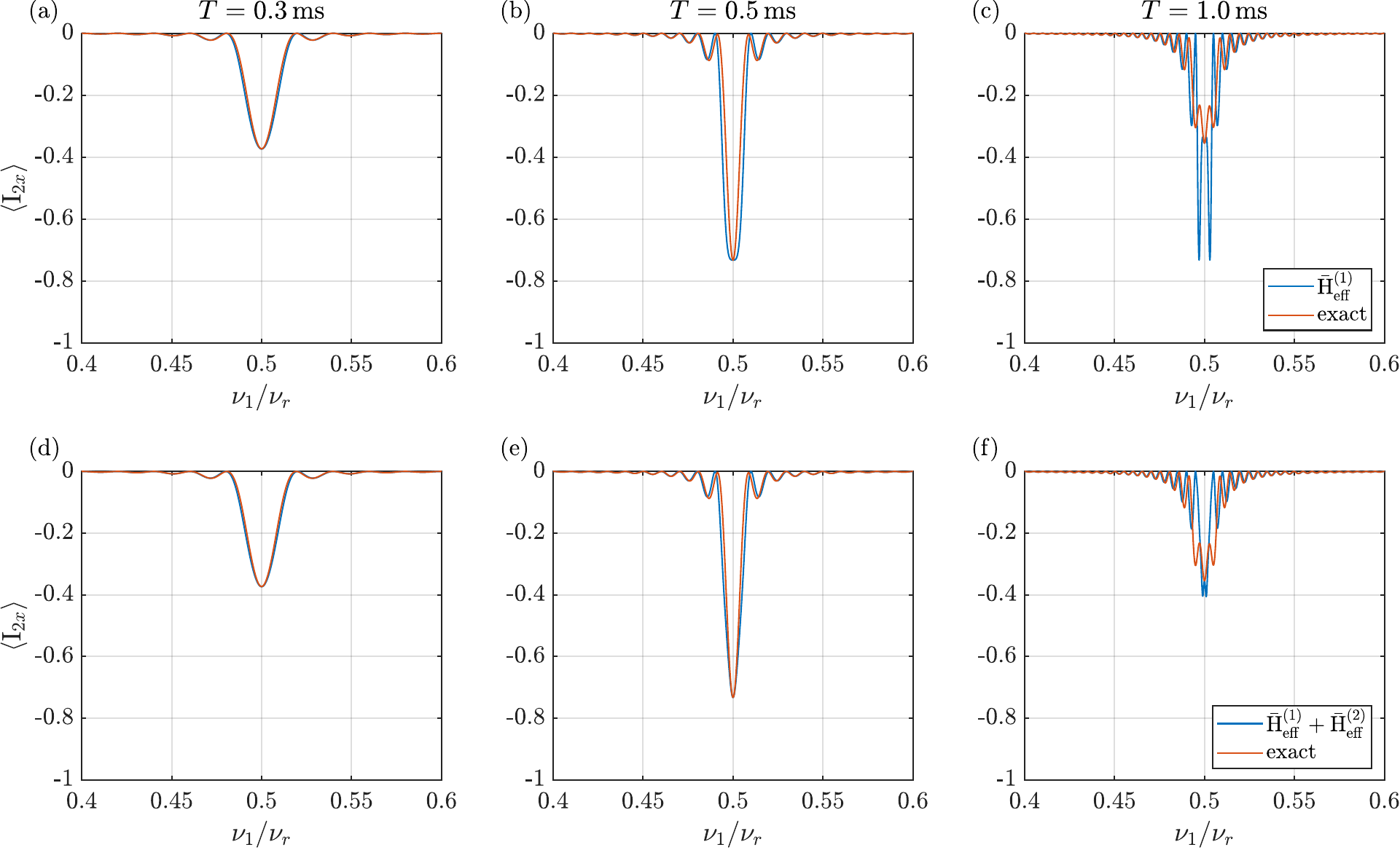}
  \caption{Comparison between numerical time-slicing (exact) and effective Hamiltonian simulation of the polarization-transfer intensity $\langle \II_{2x} \rangle$ near the HORROR resonance condition $\nu_1/\nu_r = 0.5$ for different durations $T$.
    Different crystallites were simulated with 300 spherical-powder grid orientations for powder averaging \cite{Suzukawa:1973wk}. For a standard homonuclear one-bond CC coupling, the dipolar coupling strength was set to $\delta_\mathrm{CC}/(2\pi) = -4.5\,$kHz.
   Subfigures (a)-(c) show the comparison between the recoupling intensity $\langle \II_{2x} \rangle$ obtained from the exact and first-order effective Hamiltonian simulation. 
  For durations, longer than the duration needed to achieve maximal transfer intensity ((b) and (c)), the agreement with the exact calculation decreases with increasing duration, especially near the resonance condition.
    Subfigures (d)-(f) show the comparison between the recoupling intensity $\langle \II_{2x} \rangle$ obtained from exact and up to second-order effective Hamiltonian simulation. For times up $0.5\,$ms the second-order effective Hamiltonian has almost zero contribution to the transfer intensity $\langle \II_{2x} \rangle$.
    For large durations, greatly exceeding the time needed to achieve the maximal recoupling intensity ((c) and (f)), including the second-order effective Hamiltonian significantly the agreement with exact simulations compared to calculations using only the first-order effective Hamiltonian.\label{fig:HORROR}}
\end{figure*}

Fig.~\ref{fig:HORROR} shows the comparison between numerical time-slicing (exact) simulation and effective Hamiltonian simulation of the transfer intensity $ \II_{1x}(0) \longrightarrow \II_{2x} (T)$ (in the laboratory frame) for a homonuclear two-spin system as a function of resonance mismatch $\nu_1/\nu_\rr$ for the durations $T = 0.3\,$ms, $T=0.5\,$ms and $T=1\,$ms. The MAS frequency $\nu_\rr = 100\,$kHz has been kept constant, whereas the rf-field amplitude $\nu_1$ was varied. The initial density matrix was set to $\rho_0=\II_{1x}$ (along the rf field) and detecting  $\II_{2x}$ after the duration $T$. Different crystallites were simulated with 300 spherical-powder grid orientations for powder averaging \cite{Suzukawa:1973wk}. The dipolar coupling strength was set to $\delta_\mathrm{CC}/(2\pi) = -4.5\,$kHz corresponding to a typical one-bond CC coupling. Due to the double-quantum nature of the polarization transfer, the transferred polarization appears with negative intensity.
For $T=0.3\,$ms the exact transfer intensity $\langle\II_{2x}\rangle$ is reproduced by the first-order effective Hamiltonian (Fig. \ref{fig:HORROR}a), with no visible contribution from the second-order effective Hamiltonian (Fig. \ref{fig:HORROR}d).
$T=0.4\,$ms slightly exceeds the time needed to achieve maximal recoupling, leading to minor deviations between the recoupling intensity derived by exact and first-order effective Hamiltonian (Fig. \ref{fig:HORROR}b).  However, the simulation up to the second-order effective Hamiltonian (Fig. \ref{fig:HORROR}e) accurately reproduces the transfer intensity. Note, that the time needed to achieve maximal recoupling is approximately proportional to the inverse of the dipolar coupling strength.
$T = 1\,$ms surpasses the time needed to achieve maximum recoupling intensity, leading to deviations between exact and first-order effective Hamiltonian simulation close to the resonance condition (Fig. \ref{fig:HORROR}c).
Using the first and second-order effective Hamiltonian leads to more accurate simulation (Fig. \ref{fig:HORROR}f) of the transfer intensity for $T = 1\,$ms than simulations using only the first-order Hamiltonian.
The recoupling bandwidth is in good approximation proportional to $T^{-1}$, hence decreasing the duration $T$ leads to a larger bandwidth of the transfer intensity. 
Note, that on the resonance condition, the first-order effective Hamiltonian always reproduces the transfer intensity for all mixing times.

%
%
\subsubsection{Rotary-Resonance Recoupling}
\begin{figure*}[ht]
  \centering
  \includegraphics[width=\textwidth]{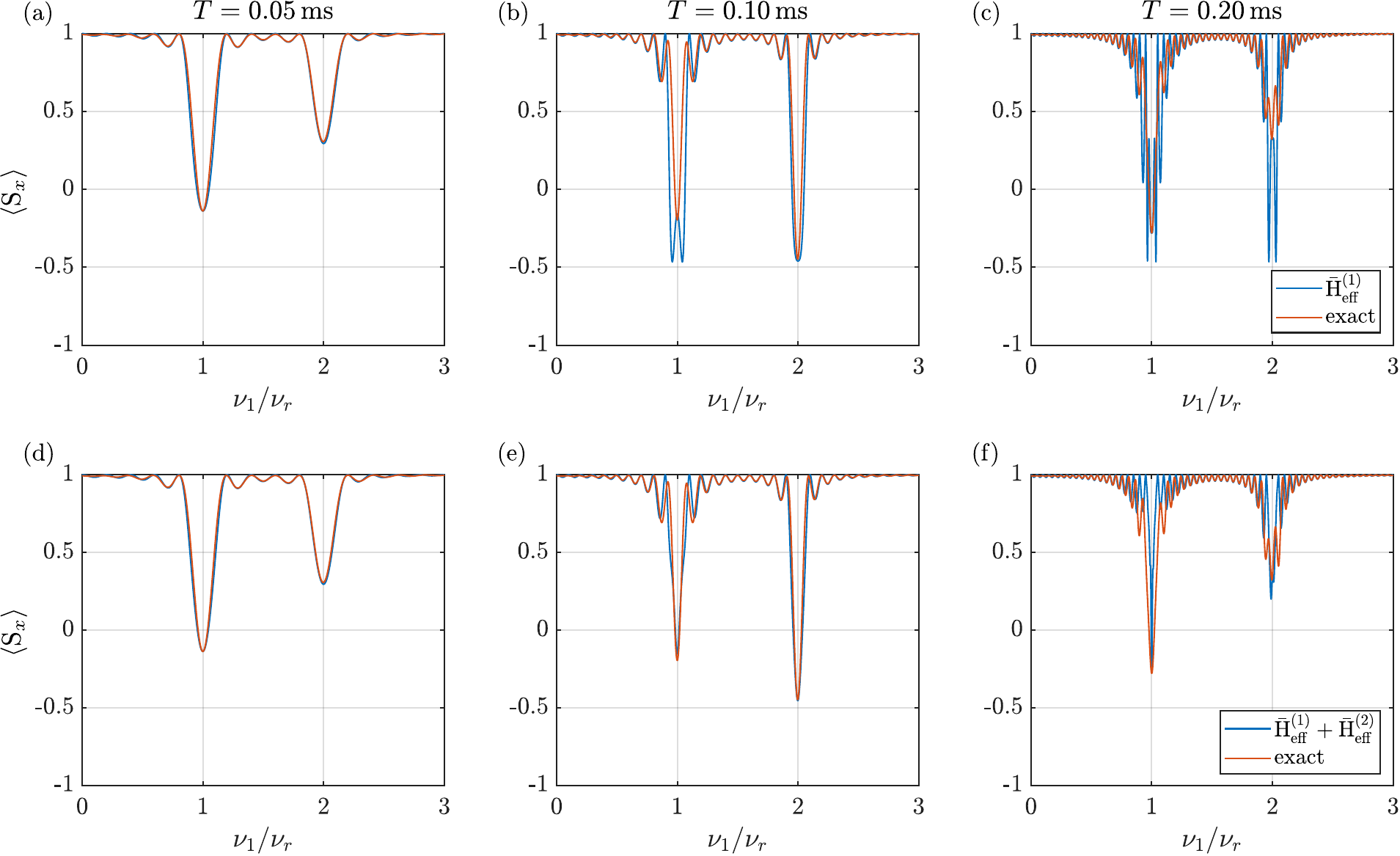}
  \caption{Comparison between exact and effective Hamiltonian simulations of the dephasing of $\SSS_{x}$ of a heteronuclear two-spin system as a function of  $\nu_1/\nu_r$ for different durations $T$.
As before, 300 spherical-powder grid orientations were used for powder averaging \cite{Suzukawa:1973wk}. The dipolar coupling strength was set to $\delta_{\mathrm{CH}}/(2\pi) = -46\,$kHz, corresponding to a standard heteronuclear one-bond CH coupling.
   Subfigures (a)-(c) illustrate the comparison between exact and first-order effective Hamiltonian simulations, while (d)-(f) include second-order corrections. The recoupling bandwidth for the rotary resonance conditions $\omega_1 = n \omega_{\rr}$ is approximately proportional to $T^{-1}$.
    For large durations, greatly exceeding the time needed to achieve the maximal dephasing ((c) and (f)), second-order corrections become significant.
    \label{fig:rotary_resonance}}
\end{figure*}

In Fig.~\ref{fig:rotary_resonance} exact and effective Hamiltonian simulations are compared for the rotary resonance conditions $\omega_1 = n \omega_\rr$ with $n\in\{1,2\}$ for a heteronuclear I-S two-spin system with irradiation on the I-spin. In such an experiment, a dephasing of the magnetization on the S-spin is observed due to the reintroduction of the heteronuclear dipolar coupling.
As in Fig.~\ref{fig:HORROR} the MAS frequency $\nu_\rr$ was kept constant at 100\,kHz while the rf-field amplitude $\nu_1$ was varied to map out the resonance conditions. The simulation parameters have been kept consistent with the previous simulation for the HORROR recoupling, except the dipolar coupling strength was set to $\delta_\mathrm{HC}/(2\pi) = -46\,$kHz corresponding to a typical heteronuclear one-bond CH coupling.
Similar to HORROR recoupling the first-order effective Hamiltonian reproduces the dephasing accurately for durations until the maximal transfer intensity is reached (Figs. \ref{fig:rotary_resonance} (a)-(c).
For durations exceeding the first maximum of the polarization transfer, higher-order corrections are necessary and become more important for increasing dephasing durations as can be seen in Figs. \ref{fig:rotary_resonance} (d)-(f).
%
%
\subsection{C-type and R-type Pulse Schemes}
Symmetry-based $\mathrm{C}N_n^\nu$ and $\mathrm{R}N_n^\nu$ pulse schemes constitute an important class of rotor-synchronized pulse schemes used for recoupling in solid-state NMR.
A substantial body of literature has been dedicated to symmetry-based pulse schemes, including consideration based on first and second-order average Hamiltonian \cite{Levit:2002, Brinkmann:2004,
Vinogradov:2004}.
$\mathrm{C}N_n^\nu$ and $\mathrm{R}N_n^\nu$ pulse schemes consist of basic elements that are synchronized with the MAS frequency. Depending on the symmetry coefficients ($N,n,\nu$) such pulse schemes lead to selective re- or decoupling of specific interactions. The coefficient $N$ represents the number of basic elements arranged within a specified number of rotor cycles denoted by $n$. 
The phase of each basic element is $\phi = 2\pi\nu/N$ in the case of $\mathrm{C}N_n^\nu$ and $\phi = \pi \nu/N$ in the case of $\mathrm{R}N_n^\nu$ pulse schemes.
The rf Hamiltonian for such phase-modulated pulse schemes can be written as
\begin{equation}
  \HH_{\rf}(t) = \omega_1 \sum_u\biggl(\SSS_{ux}\cos(\phi(t)) + \SSS_{uy}\sin(\phi(t))\biggr)~.
\end{equation}
If we only include the rf Hamiltonian in the interaction-frame transformation the spin part of the resulting Hamiltonian is periodic with the modulation frequency of the pulse scheme $\tau_\mathrm{m}$, which is the duration of the complete symmetry-based sequence.
Because the spatial part of the Hamiltonian is periodic with the MAS frequency $\omega_\rr$, we can express the resulting Hamiltonian with Fourier series in the frequencies $\omega_\mathrm{r}$ and $\omega_\mathrm{m}$:
\begin{equation}\label{eq:Bimodal-Hamiltonian}
  \HH(t)=\sum_{n,k} \widetilde{\HH}^{(n,k)} e^{i n \omega_\mathrm{r} t} e^{i k \omega_\mathrm{m} t} \Pi(t/T)~.
\end{equation}
\newpage
The Fourier coefficients of the Hamiltonian are \cite{Vinogradov:2004, Scholz:2010}:
\begin{widetext}
\begin{align}\label{eq:fourier-coefficients-C_R}
\widetilde{\HH}^{(0, k)}&=\left\{\sum_{p<q} \omega_{\mathrm{I}_p \mathrm{I}_q}^{(0)} 2 \vec{\II}_p \cdot \vec{\II}_q+\sum_{u<v} \omega_{\mathrm{S}_u \mathrm{S}_v}^{(0)} 2 \vec{\SSS}_u \cdot \vec{\SSS}_v+\sum_p \omega_{\mathrm{I}_p}^{(0)} \II_{p z}\right\} \delta_{k, 0}+\sum_u \omega_{\mathrm{S}_u}^{(0)} \sum_{s=-1}^1 a_{1, s}^{(k)} T_{1, s}^{(u)}\notag\\
&\quad +\sum_{p, u} \omega_{\II_p \SSS_u}^{(0)} 2 \II_{p z} \sum_{s=-1}^1 a_{1, s}^{(k)} T_{1, s}^{(u)} ~,\\
\widetilde{\HH}^{(n, k)}&=\left\{\sum_{p<q} \omega_{\II_p \II_q}^{(n)}\left(3 \II_{p z} \II_{q z}-\vec{\II}_p \cdot \vec{\II}_q\right)+\sum_p \omega_{\II_p}^{(n)} \II_{p z}\right\} \delta_{k, 0}+\sum_{u<v} \sqrt{6}\, \omega_{\SSS_u \SSS_v}^{(n)} \sum_{s=-2}^2 a_{2, s}^{(k)} T_{2, s}^{(u, v)} \notag\\
&\quad +\sum_u \omega_{\SSS_u}^{(n)} \sum_{s=-1}^1 a_{1, s}^{(k)} T_{1, s}^{(u)}+\sum_{p, d} \omega_{\II_p \SSS_u}^{(n)} 2 \II_{p z} \sum_{s=-1}^1 a_{1, s}^{(k)} T_{1, s}^{(u)} ~,
\end{align}
\end{widetext}
where
\begin{equation}\label{eq:a-noCS}
  a_{r, s}^{(k)} = \frac{1}{T} \int_{-T/2}^{T/2} a_{r,s}(t) e^{-i k \omega_\mathrm{m} t} \dd t~.
\end{equation}
are the Fourier coefficients of the interaction-frame transformation of the $T_{r, s}^{(p)}$ operator.
The expression for $T_{r,s}^{(p)}$ can be located in Chapter 2, Table 2.3 of the book by Mehring \cite{Mehring:1983wm}.
To obtain the effective Hamiltonian we simply insert Eq.~(\ref{eq:Bimodal-Hamiltonian}) into Eq.~(\ref{eq:first-order-effective-hamiltonian}) and Eq.~(\ref{eq:second-order-effective-hamiltonian}).
An alternative approach is to include the chemical-shift Hamiltonian in the interaction-frame transformation together with the rf Hamiltonian.
One notable advantage of this approach is that the chemical-shift terms in the Hamiltonian  Eq.~(\ref{eq:Bimodal-Hamiltonian}) are eliminated because the chemical-shift and the rf Hamiltonian are encoded in the interaction frame trajectory $a_{r,s}(t)$.
However, in this interaction frame the effective field is not zero and therefore the interaction-frame trajectory $a_{r,s}(t)$ lacks periodicity with $\omega_{\mathrm{m}}$.
Hence utilizing a Fourier series solely with $\omega_\mathrm{m}$ is not possible, but $a_{r,s}(t)$ can be expressed using the frequencies $\omega_\mathrm{m}$ and $\omega_{\eff}$, i.e.
\begin{equation}\label{eq:a-CS}
  a_{r, s}^{(k,\ell)} = \frac{1}{T} \int_{-T/2}^{T/2} a_{r,s}(t) e^{-i k \omega_\mathrm{m} t} e^{-i \ell \omega_\mathrm{eff} t} \dd t~.
\end{equation}
The Hamiltonian is therefore written in terms of three characteristic frequencies 
$\omega_\rr$, $\omega_\mathrm{m}$ and $\omega_\eff$:
\begin{equation}
  \label{eq:Trimodal-Hamiltonian}
  \widetilde{\HH}(t) = \sum_{n,k,l} \widetilde{\HH}^{(n,k,l)} e^{i (n \omega_\rr + k \omega_\mathrm{m} + \ell \omega_{\eff})t}~.
\end{equation}
As a result, the resonance conditions are $n\omega_\rr + k \omega_\mathrm{m} + \ell \omega_{\eff}$ with  $n \in \{-2,-1,..,2\}$, $k \in \mathbb{R}$ and $\ell \in \{-1,0,1\}$.
As before the first and second-order effective Hamiltonian are calculated using Eq.~(\ref{eq:first-order-effective-hamiltonian}) and Eq.~(\ref{eq:second-order-effective-hamiltonian}).
The two approaches, Eq.~(\ref{eq:Bimodal-Hamiltonian}) and Eq.~(\ref{eq:Trimodal-Hamiltonian}), lead to similar effective Hamiltonians, however including the chemical shift into the interaction frame transformation as for Eq.~(\ref{eq:Trimodal-Hamiltonian}) can lead to faster convergence of the effective Hamiltonian than Eq.~(\ref{eq:Bimodal-Hamiltonian}) as described in \cite{Chavez2022a}.
Note, that the two approaches only differ if chemical shifts are considered.

\subsubsection{C-type Pulse Schemes}
\label{sec:C-type}
\begin{figure*}[ht]
  \centering
  \includegraphics[width=\textwidth]{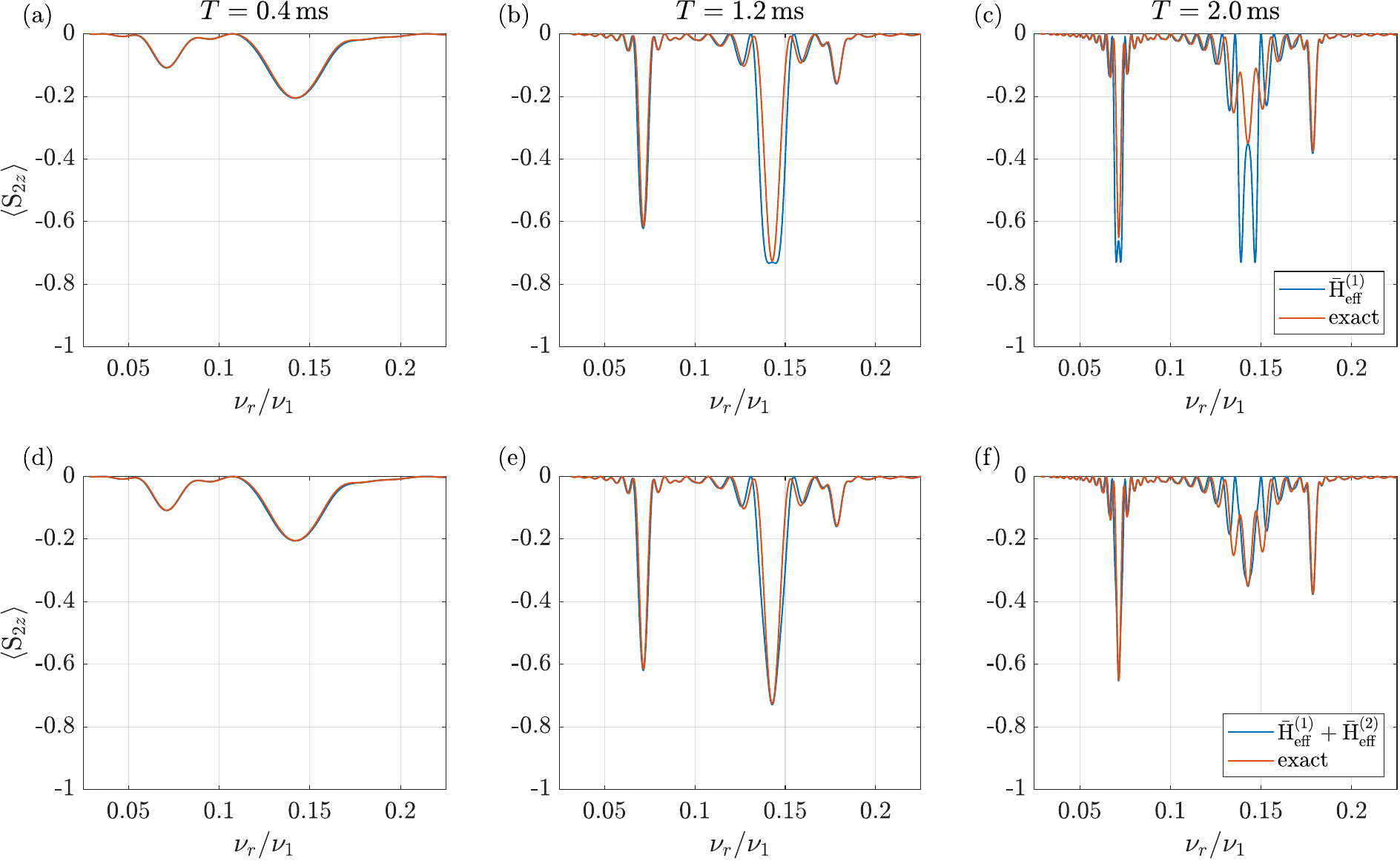}
  \caption{Transferred magnetization $\langle \SSS_{2z} \rangle$ under $\mathrm{C}$-type pulse schemes obtained from exact calculations and effective Hamiltonian calculations.
The three visible peaks correspond to $\mathrm{C}7^1_4$, $\mathrm{C}7^1_2$ and $\mathrm{C}28^4_{5}$, reading form left to right. 
Subfigures (a)-(c) show the comparison between exact and first-order effective Hamiltonian simulations, for different durations $T$. The comparison between exact and up to second-order effective Hamiltonian simulation is depicted in (d)-(f). The rf amplitude $\nu_1=70\,$kHz has been kept constant whereas the MAS frequency $\nu_r$ was varied. As in the previous examples, different crystallites with 300 spherical powder grid orientations were simulated for powder averaging \cite{Suzukawa:1973wk}. The dipolar coupling strength was set to  $\delta_\mathrm{CC}/(2\pi) = -4.5\,$kHz.\label{fig:C7_powder}}
\end{figure*}

\begin{figure*}[ht]
  \centering
  \includegraphics[width=\textwidth]{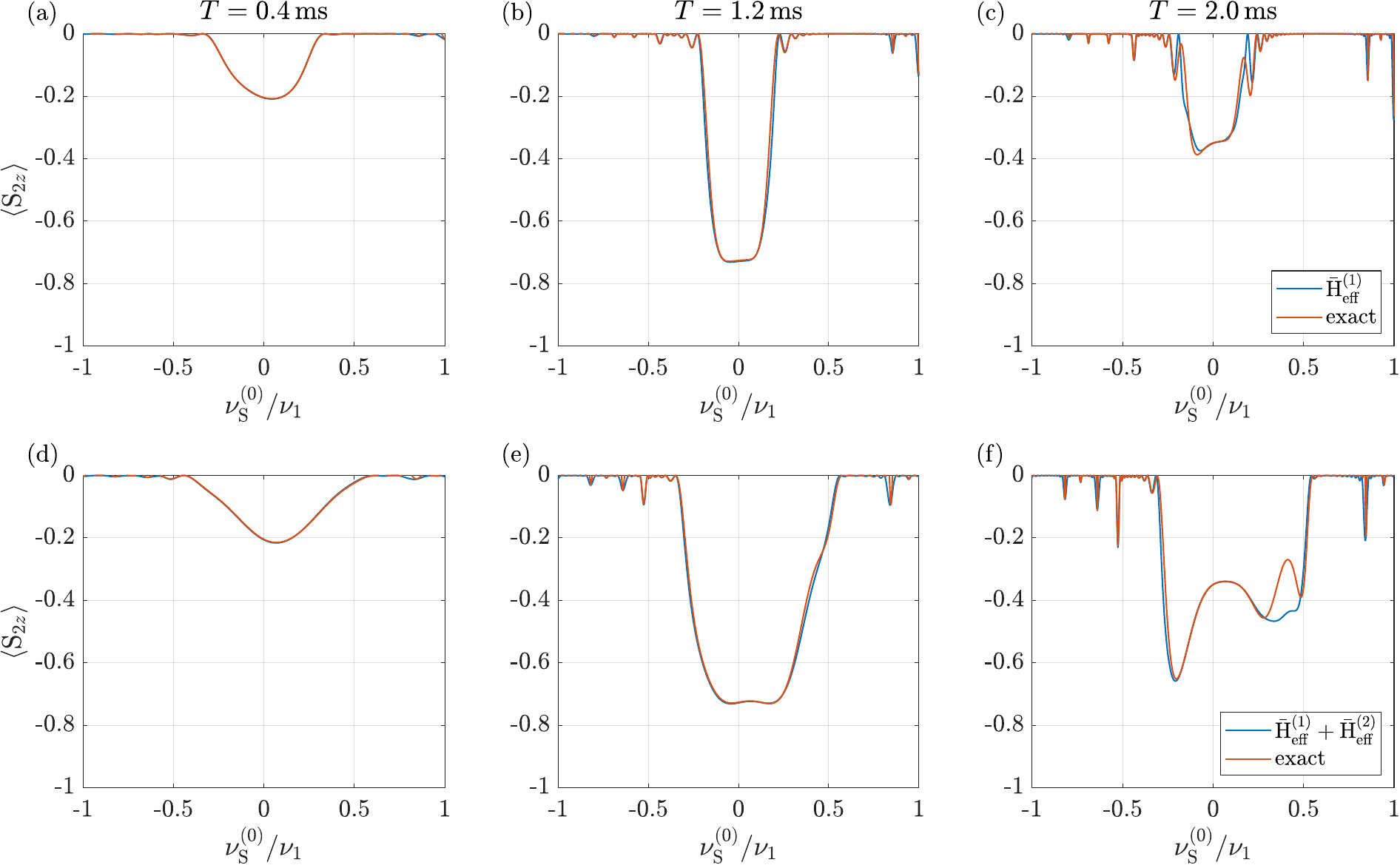}
  \caption{Recoupling intensity $\langle\SSS_{2z}\rangle$ for the $\mathrm{C}7^1_2$ and POST-$\mathrm{C}7^1_2$ pulse scheme obtained from exact calculations and effective Hamiltonian calculations up to second order as a function of $\nu^{(0)}_{\SSS}/\nu_1$.
Subfigures (a)-(c) correspond to $\mathrm{C}7^1_2$ leading to a recoupling bandwidth of approximately $0.3 \nu_1= 21\,$kHz.
In contrast, subfigures (d)-(f) show the performance of the POST-$\mathrm{C}7^1_2$ pulse scheme, which exhibits a wider recoupling bandwidth of approx. $0.8 \nu_1= 56\,$kHz. The simulation parameters are consistent with Fig.~\ref{fig:C7_powder}.\label{fig:C7_CS}}
\end{figure*}

Probably one of the most popular and historic $\mathrm{C}$-type pulse schemes is $\mathrm{C}7_n^1$, where seven basic building blocks, comprising $(2\pi)_{\phi},(2\pi)_{\phi+\pi}$, are applied within $n$ rotor periods.
Fig.~\ref{fig:C7_powder} shows the efficiency of the polarization transfer $\SSS_{1z} \rightarrow \SSS_{2z}$ by plotting $\langle \SSS_{2z}\rangle$ comparing exact numerical simulation and the time evolution under a first- and second-order effective Hamiltonian
for $\mathrm{C}7_n^1$ with different mixing times $T$.  A homonuclear two-spin system was used with a dipolar coupling strength of $\delta_\mathrm{CC}/(2\pi) = -4.5\,$kHz corresponding to a typical one-bond CC coupling. As before, 300 spherical powder grid orientations were simulated for powder averaging \cite{Suzukawa:1973wk}.
The rf amplitude $\nu_1=70\,$kHz has been kept constant whereas the MAS frequency $\nu_r$ was varied.
\\
The three areas of high polarization transfer in Fig.~\ref{fig:C7_powder} correspond to the symmetry sequences (from left to right) $\mathrm{C}7^1_4$, $\mathrm{C}7^1_2$ and  $\mathrm{C}28^4_{5}$.
The peak associated with $\mathrm{C}7^1_2$ achieves maximum recoupling intensity at the fastest rate and exhibits the broadest width. Subsequently, the $\mathrm{C}7^1_4$ peak, being the second broadest, follows, while the narrowest peak corresponds to $\mathrm{C}28^4_{5}$ reflecting different scaling factors for the dipolar coupling.
As in the previous examples, the exact recoupling intensity is replicated using only the first-order effective Hamiltonian for durations $T$ up to the time required to obtain the maximum transfer intensity.
For durations exceeding the maximum transfer intensity, the second-order correction becomes important.  Corrections above second-order are only important when the duration $T$ is much bigger than the time needed to achieve maximum transfer.
An important property of recoupling pulse schemes is the sensitivity to the chemical-shift offset of the irradiated spin $\nu_{\mathrm{S}}^{(0)}$. Ideally, these schemes should exhibit effective recoupling across a broad range of chemical-shift offsets.
To extend this range the basic element of the $\mathrm{C}N_n^\nu$ pulse schemes $(2\pi)_{\phi},(2\pi)_{\phi+\pi}$ is substituted with the composite pulse $(\pi/2)_\phi,(2\pi)_{\phi+\pi},(3\pi/2)_{\phi}$ \cite{Hohwy:1998vv}.
These modified $\mathrm{C}$ pulse schemes, termed POST-$\mathrm{C}N_n^\nu$, are commonly used in experimental implementations due to their better offset compensation.
Figure~\ref{fig:C7_CS} shows the transferred polarization as a function of the chemical-shift offset $\nu^{(0)}_{\SSS}$ for the $\mathrm{C}7^1_2$ and POST-$\mathrm{C}7^1_2$ pulse scheme. As expected, the POST-$\mathrm{C}7^1_2$ pulse scheme exhibits a better chemical-shift offset bandwidth of approx. $0.8 \nu_1= 56\,$kHz, compared to the standard $\mathrm{C}7^1_2$ pulse scheme with a bandwidth of approx. $0.3 \nu_1= 21\,$kHz. The simulation parameters, including the spin system parameters and rf amplitude, remain consistent with those utilized in Fig.~\ref{fig:C7_powder}. The effective Hamiltonians were calculated using Eq.~(\ref{eq:Trimodal-Hamiltonian}), thus in the interaction frame including the rf and the chemical-shift Hamiltonian.

\subsubsection{R-type Pulse Schemes}
\label{sec:R-type}
\begin{figure*}[ht]
  \centering
  \includegraphics[width=\textwidth]{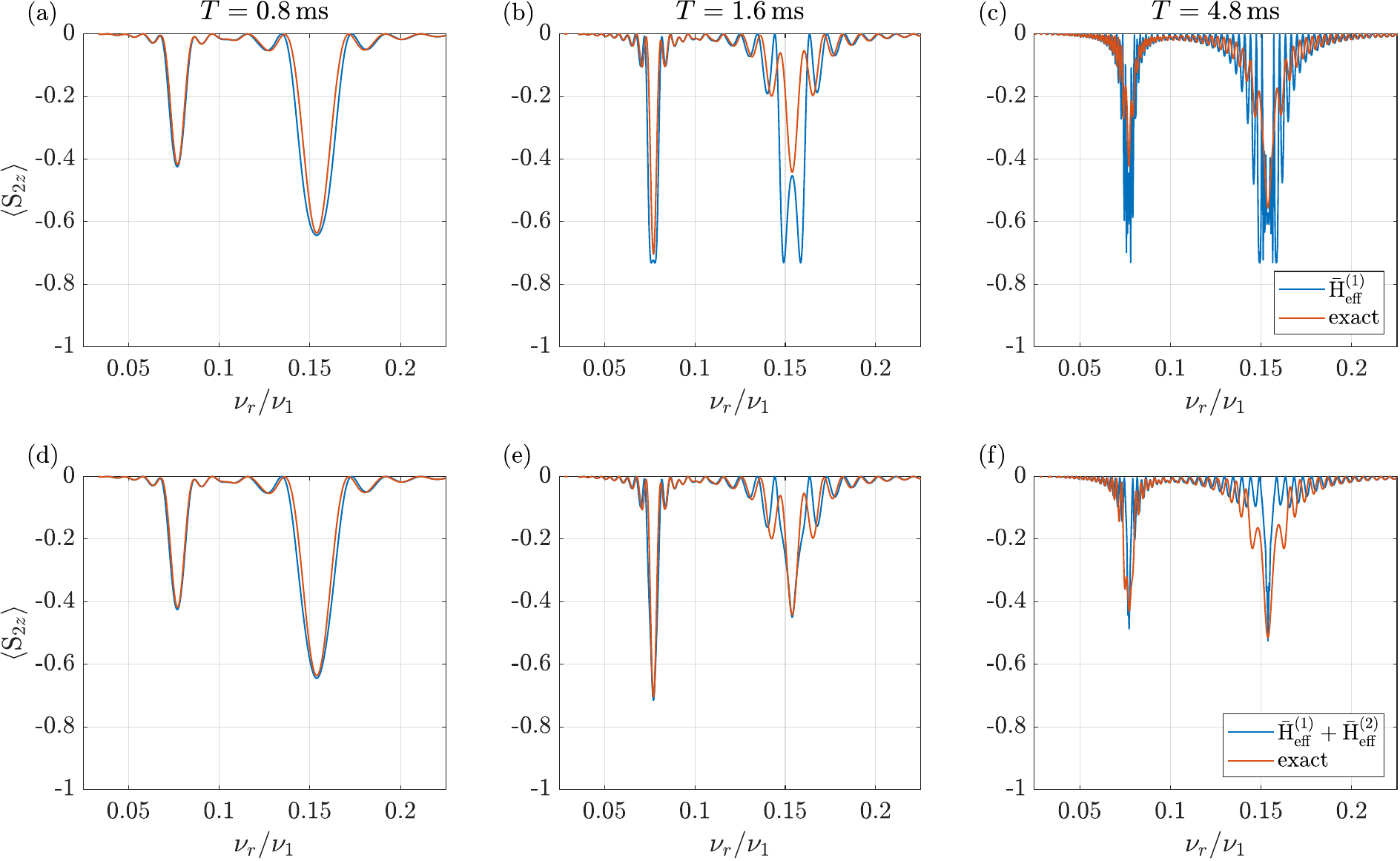}
  \caption{Recoupling intensity $\langle \SSS_{2z} \rangle$ under $\mathrm{R}$-type pulse schemes obtained from exact calculations and effective Hamiltonian calculations.
The left peak corresponds to the $\mathrm{R}26^{11}_{8}$ whereas the right peak corresponds to $\mathrm{R}26^{11}_{4}$. 
Subfigures (a)-(c) show the comparison between exact and first-order effective Hamiltonian simulations, for different durations $T$. The comparison between exact and up to second-order effective Hamiltonian simulation is represented in subfigure (d)-(f). The rf amplitude $\nu_1=70\,$kHz has been kept constant whereas the MAS frequency $\nu_r$ was varied. The initial density matrix was set to $\II_{1z}$ and after the duration $T$ $\II_{2z}$ was detected. \label{fig:R26_powder}}
\end{figure*}

\begin{figure*}[ht]
  \centering
  \includegraphics[width=\textwidth]{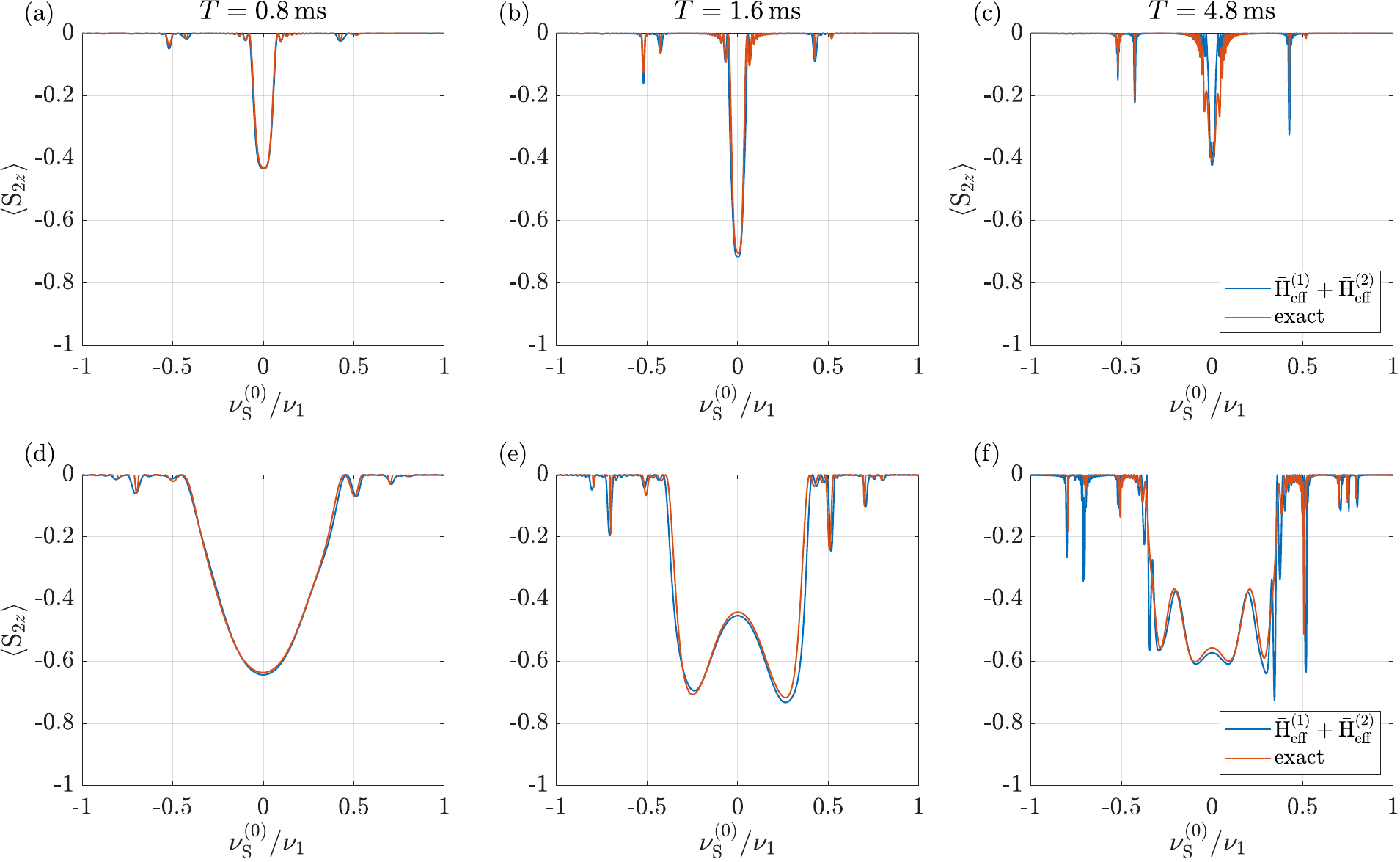}
  \caption{Recoupling intensity $\langle\SSS_{2z}\rangle$ for the $\mathrm{R}26^{11}_4$ pulse scheme obtained from exact calculations and effective Hamiltonian calculations up to second order as a function of $\nu^{(0)}_{\SSS}/\nu_1$ for $\nu_1 = 65\,$kHz.
 Subfigures (a)-(c) correspond to the implementation $\mathrm{R}26^{11}_4$ without composite pulses leading to a recoupling bandwidth of approximately $0.1 \nu_1= 6.5\,$kHz.
For subfigures (d)-(f), $\mathrm{R}26^{11}_4$ was implemented using composite pulses increasing the recoupling bandwidth significantly to approximately $0.8 \nu_1= 52\,$kHz.
The simulation parameters have been kept consistent with the parameters used for Fig.~\ref{fig:C7_CS}.\label{fig:R26_powder_CS}}
\end{figure*}

Typical implementation of the basic element of $\mathrm{R}$-type pulse schemes are either $(\pi)_{\phi},(\pi)_{-\phi}$, or using composite pulse $(\pi/2)_\phi(3\pi/2)_{\phi+\pi}$ for the basic $\pi$ rotation.
Implementation using composite pulses increases the chemical-shift compensation and, therefore, the recoupling bandwidth of the sequence while at the same time requiring a higher rf-field amplitude at the same spinning frequency. 
Fig.~\ref{fig:R26_powder} shows the transfer intensity $\langle \SSS_{2z}\rangle$ comparing exact and first and second-order effective Hamiltonian simulations as a function of $\nu_1/\nu_r$ for $\mathrm{R}26_8^{11}$ (left peak) and $\mathrm{R}26_4^{11}$ (right peak) for different $T$.
The parameters for the simulations have been kept consistent with those used for the previous simulation of the C-type pulse schemes.
As in the previous cases, the first-order Hamiltonian is sufficient to reproduce the transfer intensity $\langle S_{2z} \rangle$ for durations up to the point where maximal transfer intensity is attained. For longer durations, second-order corrections become necessary. 
Fig.~\ref{fig:R26_powder_CS} compares the transfer intensity for the $\mathrm{R}26^{11}_4$ pulse scheme using the basic element $(\pi)_{\phi},(\pi)_{-\phi}$ ((a)-(c)) and $(\pi/2)_\phi(3\pi/2)_{\phi+\pi}$ ((d)-(e)) as a function of $\nu^{(0)}_{\SSS}/\nu_1$ for $\nu_1 = 65\,$kHz. 
Using composite pulses greatly improves the recoupling bandwidth from approximately $0.1 \nu_1= 6.5\,$kHz to $0.8 \nu_1= 52\,$kHz.

%
%
\section{Conclusion}
Effective Hamiltonians serve as indispensable tools for comprehending and developing solid-state NMR experiments. Traditional Floquet theory has proven particularly efficacious in experiments involving multiple modulation frequencies, often entailing MAS and employing multiple rf pulses on one or more spin species.

The prerequisite of a periodic Hamiltonian, inherent in traditional Floquet theory, can, however, be restrictive. Continuous Floquet theory represents an extension of its traditional counterpart, eliminating the necessity of a periodic Hamiltonian. This extension enables the analysis of both non-resonant and near-resonant experiments, affording a more extensive back-calculation of the effective Hamiltonian from experimental parameters.

In Continuous Floquet theory, the definition of the effective Hamiltonians uses a single frequency to describe multiple time dependencies. This feature, while enhancing the scope of analysis, may complicate the interpretation of effective Hamiltonians. Furthermore, the definition of the second-order effective Hamiltonian as a Cauchy principal value introduces certain inconveniences, potentially leading to decreased computational efficiency. One disadvantage of Continuous Floquet theory is the fact that the effective Hamiltonian depends on the length of the sequence unless it is calculated on a resonance condition where it is the same as the classical effective Floquet Hamiltonian. This requires a recalculation of the convolution of the Fourier coefficients with the sinc function if the mixing time is changed.

The effective Hamiltonians in this paper, resolve these shortcomings and closely align with traditional Floquet theory, ensuring ease and flexibility in their application to NMR experiments with multiple modulation frequencies. Notably, Fourier coefficients obtained through analysis within the traditional Floquet theory framework can be repurposed seamlessly, establishing a unified foundation for computational methodologies.

In the later part of this manuscript, we focus on applying the presented framework, emphasizing earlier discussed aspects.
Utilizing Fourier coefficients from literature calculated within traditional Floquet theory, we investigated the near-resonance behavior of various recoupling pulse schemes. Our analysis emphasized that the overall duration is the most crucial parameter for recoupling bandwidth in these experiments.

\section*{Supplementary Material}
Simulation of single crystallites using different dipolar coupling strengths for the discussed pulse schemes, as well as the derivation of the first and second effective Hamiltonian, can be found in the Supplementary Information.

\section*{Acknowledgments}
This research was supported by the ETH Zürich and the Schweizerischer Nationalfonds zur Förderung der Wissenschaftlichen Forschung (grant nos. 200020\_188988 and 200020\_219375).

\section*{Author Declarations}

\subsection*{Conflict of Interest}
The authors have no conflicts to disclose.

\subsection*{Author Contributions}
Conceptualization (MC,ME); Formal analysis (MC);
Investigation (MC,ME); Methodology (MC); Software (MC,ME); Writing – original draft (MC); Writing – review \& editing (MC,ME).

\section*{Data Availability}
The data that support the findings of this study are available on \href{https://doi.org/10.3929/ethz-b-000668159}{https://doi.org/10.3929/ethz-b-000668159}.
%
%
\nocite{*}
\bibliography{bibiliography}

\end{document}


\maketitle
%
%
\section{Derivation of the first and second-order effective Hamiltonian}
To derive the first and second-order effective Hamiltonian, 
we first have to calculate the frequency-domain Hamiltonian $\widehat{\HH}(\Omega)$, defined as the element-wise Fourier transform of the original time-dependent Hamiltonian
\begin{align}\label{eq:fd1-hamiltonian}
  \widehat{\HH}(\Omega) = \int\limits_{-\infty}^\infty\HH(t)e^{-i \Omega t}\dd t~.
\end{align}
Inserting the Hamiltonian
\begin{align}
    \HH(t)  =  \widetilde{\HH}(t) \Pi(t/T) =  \sum_{\mathbf{n}} \widetilde{\HH}^{(\mathbf{n})}   e^{i \omega_\mathbf{n} t} \Pi(t/T)
\end{align}
into the definition of the frequency-domain Hamiltonian $\widehat{\HH}(\Omega)$ (Eq.~(\ref{eq:fd1-hamiltonian})) leads to
\begin{align}
  \widehat{\HH}(\Omega) &=\sum_{\mathbf{n}} \widetilde{\HH}^{(\mathbf{n})}  \int\limits_{-\infty}^\infty    e^{i (\omega_\mathbf{n}-\Omega) t} \Pi(t/T) \dd t \notag\\
                        &=\sum_{\mathbf{n}} \widetilde{\HH}^{(\mathbf{n})}  \int\limits_{-\frac{T}{2}}^{\frac{T}{2}} e^{i (\omega_\mathbf{n}-\Omega) t} \dd t \notag\\
                        &= T \sum_{\mathbf{n}} \HH^{(\mathbf{n})} \mathrm{sinc}\left(\frac{(\omega_\mathbf{n}-\Omega )T}{2}\right)~.\label{eq:fd-hamiltonian}
\end{align}
\subsection{First-order effective Hamiltonian}
\label{sec:first-order-effect}
The first-order effective Hamiltonian is calculated directly from the frequency-domain Hamiltonian, resulting in
\begin{align}
  \label{eq:first-order-effective-hamiltonian}
  \bar{\HH}^{(1)}= \frac{1}{T} \widehat{\HH}(0)
  &= \sum_{\mathbf{n}} \widetilde \HH^{(\mathbf{n})} h^{(1)}_{\mathbf{n}}~,
\end{align}
with
\begin{align}
  h^{(1)}_{\mathbf{n}} = \mathrm{sinc}\left(\frac{(\omega_\mathbf{n}-\Omega )T}{2}\right)~.
\end{align}
\subsection{Second-order effective Hamiltonian}
\label{sec:second-order-effect}
Inserting the frequency-domain Hamiltonian $\widehat{\HH}(\Omega)$ (Eq.~(\ref{eq:fd-hamiltonian}))  into the definition of the second-order effective Hamiltonian leads to
\begin{align}
\bar{\HH}^{(2)}
&= -\frac{1}{2 T} \, PV
\int 
\frac{[\widehat{\HH}(\Omega)
	  ,\widehat{\HH}(-\Omega)]}
       {\Omega}\dd\Omega \notag\\
&=  -\frac{1}{2 T}
\sum_{n,m} [\HH^{(n)}
  ,\HH^{(m)}] 
  T^2 PV\int
  \frac{\text{sinc}\left(\frac{T}{2} (\omega_{\mathbf{n}}-\Omega)\right) \text{sinc}\left(\frac{T}{2} (\omega_{\mathbf{m}}+\Omega )\right)}{\Omega}
  \dd\Omega \notag\\
& = -\frac{1}{2}
  \sum_{n,m}[\HH^{(n)},\HH^{(m)}] 
  \frac{2}{T}\frac{ \omega_{\mathbf{m}} \sin \left(\frac{\omega_{\mathbf{n}} T}{2}\right) \cos \left(\frac{\omega_{\mathbf{m}} T}{2}\right)-\omega_{\mathbf{n}} \cos \left(\frac{\omega_{\mathbf{n}} T}{2}\right) \sin \left(\frac{\omega_{\mathbf{m}} T}{2}\right)}{\omega_{\mathbf{n}} \omega_{\mathbf{m}} (\omega_{\mathbf{n}}+\omega_{\mathbf{m}})}~.
\label{eq:time-decomposition-second-order}
\end{align}
This results in the second-order effective Hamiltonian
\begin{align}
  \label{eq:second-order-effective-hamiltonian}
    \bar{\HH}^{(2)}
  &= -\frac{1}{2}\sum_{\mathbf{n},\mathbf{m}}[\widetilde{\HH}^{(\mathbf{n})},\widetilde{\HH}^{(\mathbf{m})}]
    \,h^{(2)}_{\mathbf{n},\mathbf{m}}(T)
\end{align}
with
\begin{equation}
  \label{eq:basis-function2}
  h^{(2)}_{\mathbf{n},\mathbf{m}}(T) = \frac{2}{T} \frac{ \omega_\mathbf{m} \sin \left(\frac{\omega_\mathbf{n} T}{2}\right) \cos \left(\frac{\omega_\mathbf{m} T}{2}\right)-\omega_\mathbf{n} \cos \left(\frac{\omega_\mathbf{n} T}{2}\right) \sin \left(\frac{\omega_\mathbf{m} T}{2}\right)}
  {\omega_\mathbf{n} \omega_\mathbf{m} (\omega_\mathbf{n}+\omega_\mathbf{m})}
  ~.
\end{equation}

%
%

\section{Single crystallite simulation}
\label{sec:single-crystal}
The simulations shown in the main text use different crystallites with 300 spherical-powder grid orientations for powder averaging. In the following simulations of a single crystallite are shown.

%
%
\subsection{HORROR}
\begin{figure}[H] 
  \centering
  \includegraphics[width=1\textwidth]{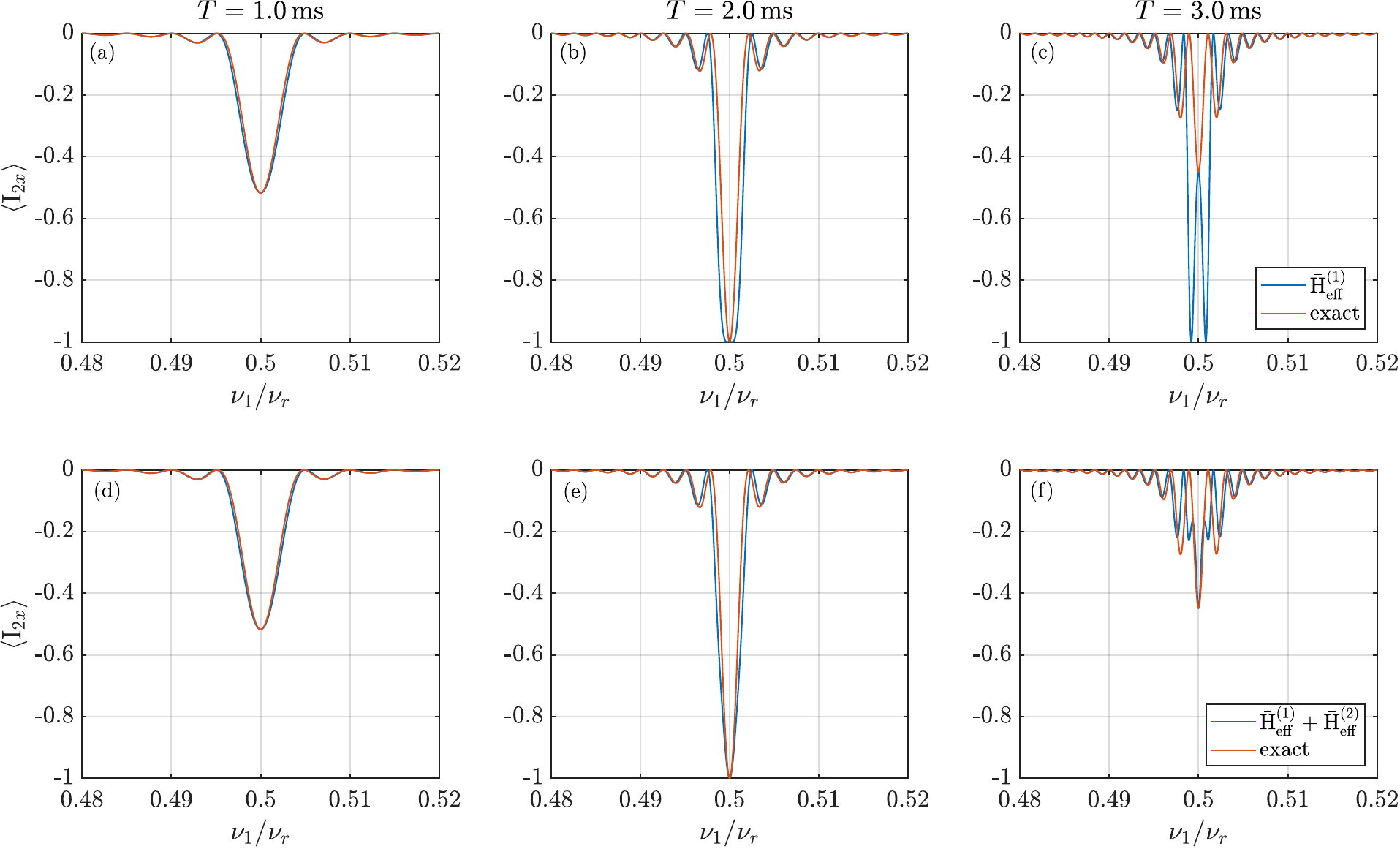}
  \caption{
    Comparison between exact and effective Hamiltonian simulations of the transfer intensity $\langle \II_{2x} \rangle$ for single crystal orientation near the HORROR resonance condition $\nu_1/\nu_r = 0.5$ for different durations $T$.
    (a)-(c) Comparison between the recoupling intensity $\langle \II_{2x} \rangle$ obtained from the exact and first-order effective Hamiltonian simulation. 
    For large durations, which are longer than the duration needed to achieve maximal transfer intensity, the agreement with the exact calculation decreases with increasing duration, especially near the resonance condition. (d)-(f) Comparison between the recoupling intensity $\langle \II_{2x} \rangle$ obtained from exact and up to second-order effective Hamiltonian simulation. For times up $2\,$ms the second-order effective Hamiltonian has almost zero contribution to the transfer intensity $\langle \II_{2x} \rangle$.
    For large durations, greatly exceeding the time needed to achieve the maximal recoupling intensity, including the second-order effective Hamiltonian significantly the agreement with exact simulations compared to calculations using only the first-order effective Hamiltonian.
  \label{fig:HORROR}}
\end{figure}

Fig.~\ref{fig:HORROR} shows the transfer intensity for a single crystal with the Euler angles $(\alpha, \beta, \gamma) = (44.4^\circ, 0.6^\circ, 73.2^\circ)
$ near the HORROR resonance condition $\nu_1/\nu_r = 0.5$ for different durations $T$. In contrast to the main text, the dipolar coupling strength was set to $\delta_{CC}/(2\pi) = -2.25\,$kHz instead of $\delta_{CC}/(2\pi) = -4.5\,$kHz. Note that maximal transfer intensity is therefore reached in twice the duration compared to the case of the stronger coupling.

\newpage
%
%
\subsection{Rotary Resonance}
\begin{figure}[H]
  \centering
  \includegraphics[width=\textwidth]{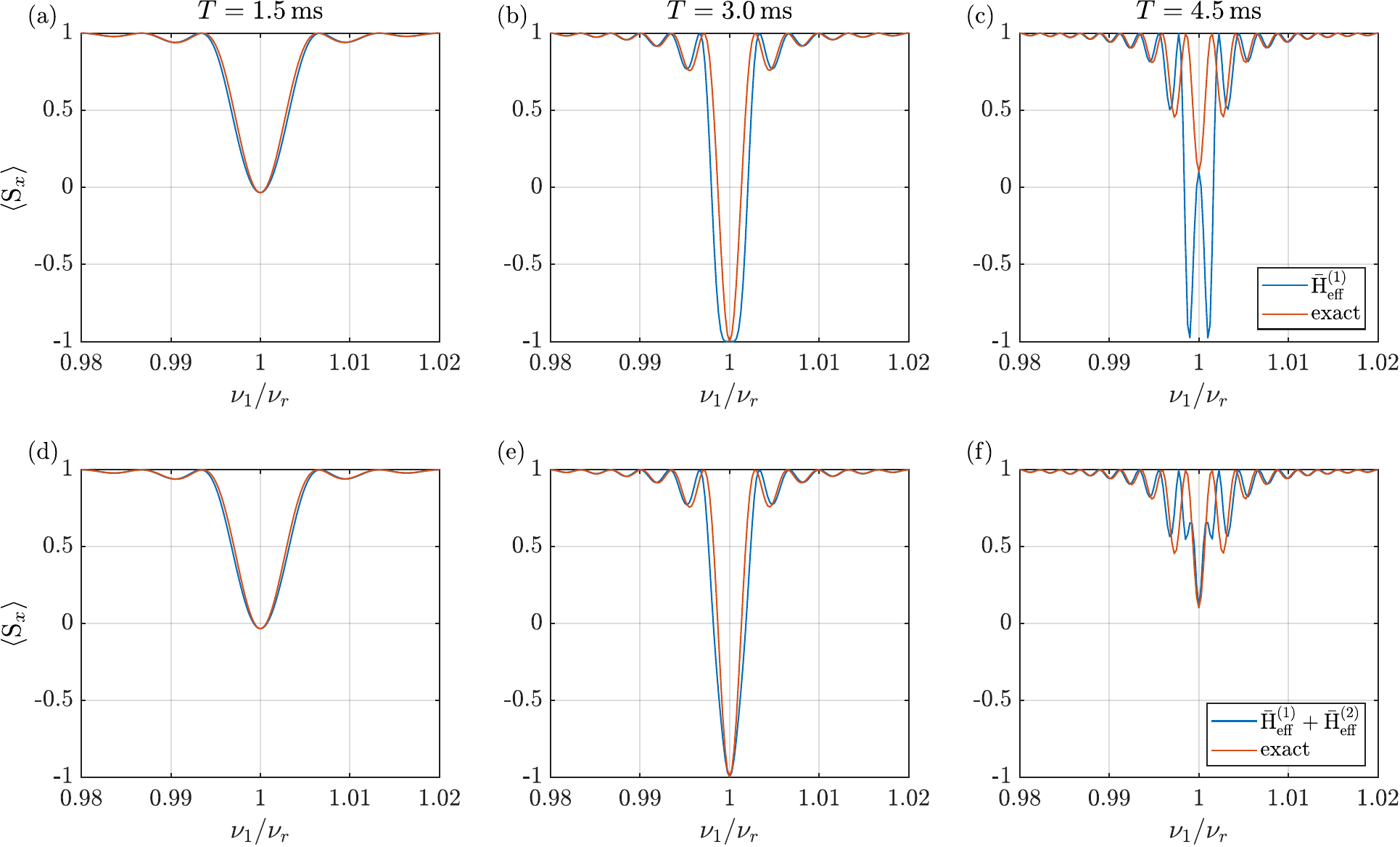}
  \caption{
    Comparison between exact and effective Hamiltonian simulations of the dephasing of $\langle \SSS_{x} \rangle$ of a heteronuclear two-spin system as a function of  $\nu_1/\nu_r$ for different durations $T$.
A single (arbitrary) orientation of a heteronuclear spin system was used with the powder angles $(\alpha, \beta, \gamma) = (44.4^\circ, 0.6^\circ, 73.2^\circ)$, where the initial density matrix was set to $\SSS_{x}$. The dipolar coupling strength $\delta_{HC}/(2\pi) = -46\,$kHz was selected to mimic a typical one-bond HC coupling. After the duration $T=1.5, 3.0$ and $4.5$\,ms $\SSS_{x}$ was detected. Subfigures (a)-(c), show the comparison between exact and first-order effective Hamiltonian simulations, for different durations $T$. Subfigures (d)-(f) illustrate the comparison between exact and up to second-order effective Hamiltonian simulation.
    \label{fig:rotary_resonance}}
\end{figure}

Fig.~\ref{fig:rotary_resonance} shows the dephasing of $S_x$ for a single crystal with the Euler angles $(\alpha, \beta, \gamma) = (44.4^\circ, 0.6^\circ, 73.2^\circ)
$ near the rotary-resonance condition $\nu_1/\nu_r = 1$ for different durations $T$.
The dipolar coupling strengths was set to $\delta_{HC}/(2\pi) = -46\,$kHz, corresponding to a typical one bond CH coupling

%
%
\subsection{$C$-type Pulse Schemes}
\begin{figure}[H]
  \centering
  \includegraphics[width=\textwidth]{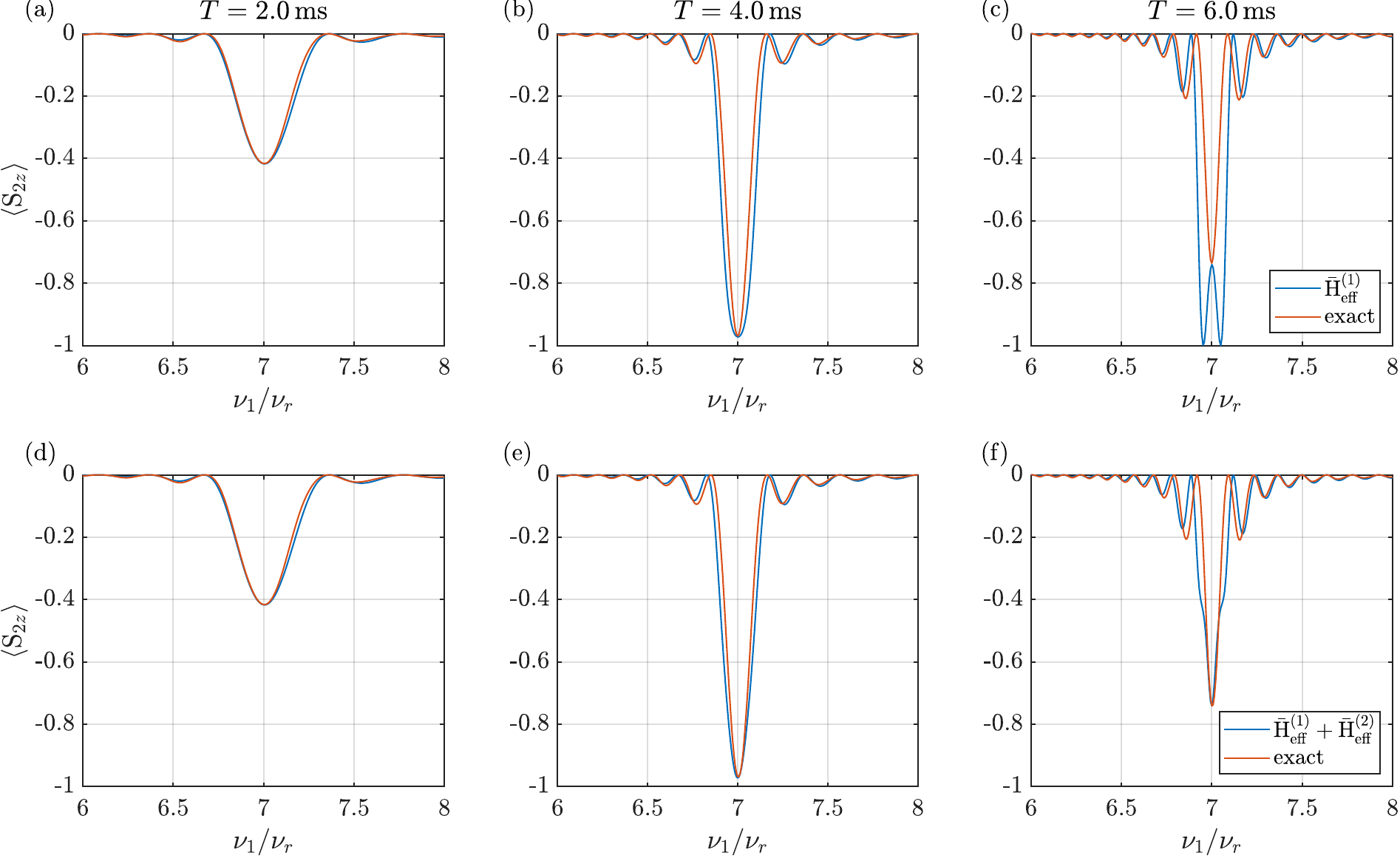}
  \caption{Transferred poarization $\langle \SSS_{2z} \rangle$ for the (POST-)$C7^1_2$ pulse scheme obtained from exact calculations and effective Hamiltonian calculations. A single crystallite orientation of a homonuclear spin system was used with the Euler angles $(\alpha, \beta, \gamma) = (44.4^\circ, 0.6^\circ, 73.2^\circ)$, where the initial density matrix was set to $\SSS_{1z}$.
The dipolar coupling strength was adjusted to $\delta/(2\pi) = -2.25\,$kHz, which led to a doubling of the time required to achieve maximum transfer intensity compared to the scenario described in the main text, where $\delta/(2\pi) = -4.5\,$kHz.
    \label{fig:postC7}}
\end{figure}
\subsection{$R$-type Pulse Schemes}
%
%
\begin{figure}[H]
  \centering
  \includegraphics[width=\textwidth]{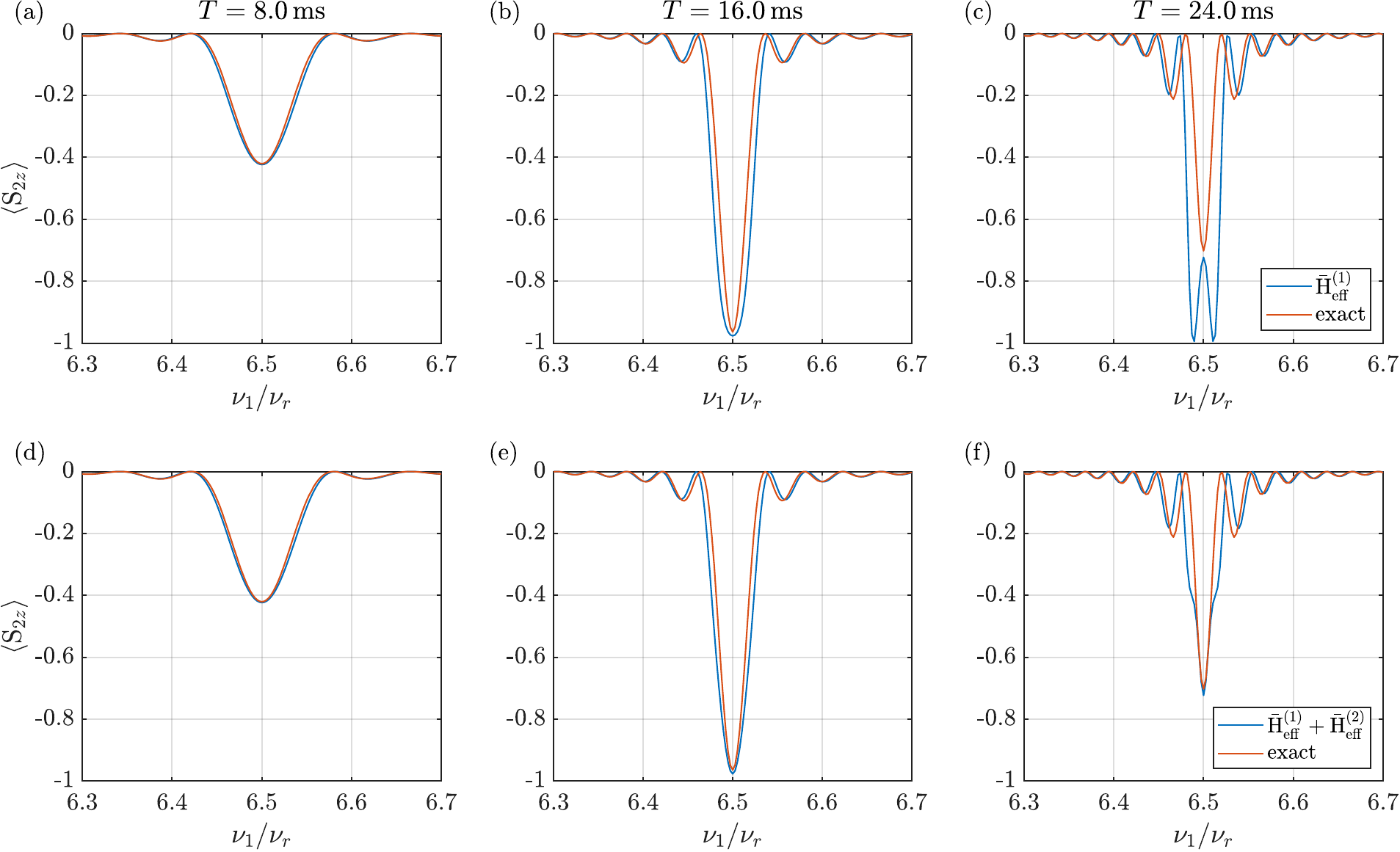}
     \caption{Transferred polarization $\langle\SSS_{2z}\rangle$ for the $R26^{11}_4$ pulse scheme obtained from exact calculations and effective Hamiltonian calculations. The dipolar coupling strength of the homonuclear two-spin system is $\delta/(2\pi) = -10.55\,$kHz. The upper row shows the comparison between exact and first-order effective Hamiltonian simulations, for different durations $T$. The lower row depicts the comparison between exact and up to second-order effective Hamiltonian simulation. A single crystallite with the Euler angles $(\alpha, \beta, \gamma) = (44.4^\circ, 0.6^\circ, 73.2^\circ)$ was used, where the initial density matrix was set to $\SSS_{1z}$. After the duration $T=8, 16, 24$\,ms (corresponding to $10, 20$ and $30$ repetition of the pulse scheme) $\SSS_{2z}$ was detected. \label{fig:R26_resonant}}
\end{figure}
%
%
\subsubsection{$R$-type Pulse Schemes with Chemical Shift}
\begin{figure}[H]
  \centering
  \includegraphics[width=\textwidth]{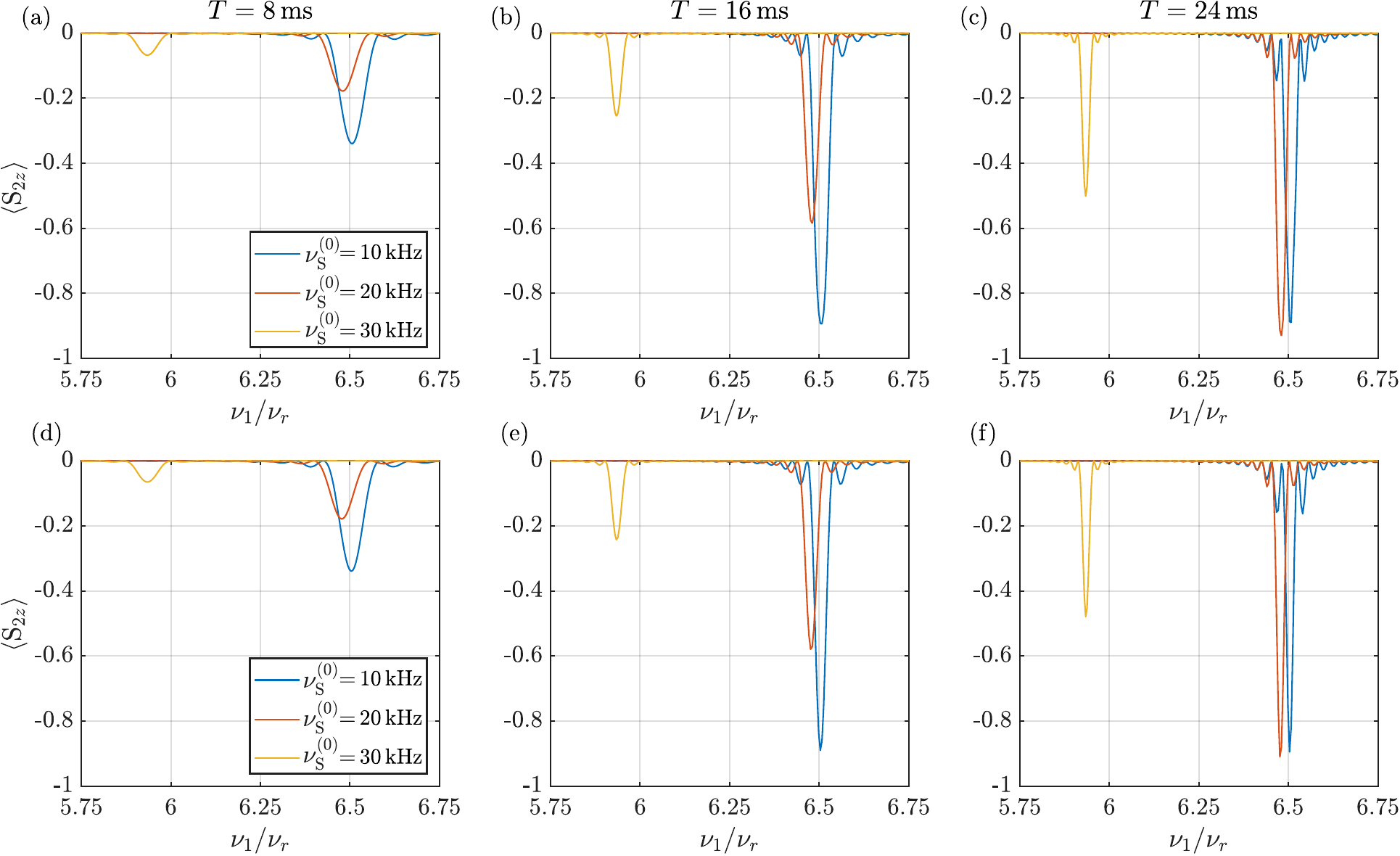}
  \caption{
    Transferred polarization $\langle \SSS_{2z} \rangle$ for the $R26^{11}_4$ pulse scheme obtained from exact ((a)-(c)) and second-order effective Hamiltonian ((d)-(f)) calculations for different chemical-shift offsets $\nu^{(0)}_{\mathrm{S}}$ and durations $T$ as a function of $\nu_1/\nu_\rr$. After the duration $T=8, 16$ and $24$\,ms $\SSS_{2z}$ was detected. As in the previous figures a single crystallite with the Euler angles $(\alpha, \beta, \gamma) = (44.4^\circ, 0.6^\circ, 73.2^\circ)$ was used. The dipolar coupling strength has been set to $\delta/(2\pi) = -10.55\,$kHz.
    \label{fig:R26_nopowder_CS}}
\end{figure}

Fig.~\ref{fig:R26_nopowder_CS} shows the transfer intensity $\langle \SSS_{2z} \rangle$ for the $R26^{11}_4$ pulse scheme for different chemical-shift offsets $\nu^{(0)}_{\mathrm{S}}$ and durations $T$ as a function of $\nu_1/\nu_\rr$. Here $\nu_1=65\,$kHz and $\nu_r$ has been varied.  As the chemical-shift offset increases, the regions of efficient transfer shift in a non-linear manner. Furthermore, the duration required to achieve maximum recoupling increases with rising chemical-shift offset, following an approximately linear trend. For the simulation a single crystallite with the Euler angles $(\alpha, \beta, \gamma) = (44.4^\circ, 0.6^\circ, 73.2^\circ)$ was used. The dipolar coupling strength has been set to $\delta/(2\pi) = -10.55\,$kHz. 
